\documentclass{eptcs}
\providecommand{\event}{TFPIE 2025}

\usepackage{subcaption}
\usepackage{caption}
\usepackage{amsmath}
\usepackage{calc}
\usepackage{enumitem}
\setlist[description]{leftmargin=\parindent,labelindent=\parindent}
\usepackage{hyperref}
\usepackage{cleveref}
\usepackage{alltt}
\usepackage{tikz}
\usepackage{float}
\floatstyle{ruled}
\restylefloat{figure}
\usepackage{balance}
\usepackage[export]{adjustbox}
\usepackage[T1]{fontenc}
\usepackage{listings}

\newcommand{\ep}{\texttt{\(\epsilon\)}}

	\lstdefinelanguage{racket} {
		morecomment=[l]{;},         
		morecomment=[s]{\#|}{|\#},  
		commentstyle={\color{darkorange}\slshape\sffamily},
		morestring=[b]",
		stringstyle=\color{red},
		literate=%
		{->}{{$\rightarrow$}}1
		{*}{{*}}1
		{lambda}{{\textcolor{blue}{\(\lambda\)}}}1
		{EMP}{{\ep{}}}1,
		classoffset=1,
		morekeywords={
			define, define-syntax, define-macro, define-datatype, define-stream, \#lang, stream-lambda
		},
		keywordstyle=\color{purple},
		classoffset=2,
		morekeywords={
			*, *list/c, +, -, /, <, </c, <=, =, >, >/c, >=, abort-current-continuation, abs, absolute-path?, acos, add1, alarm-evt, always-evt, andmap, angle, append, append*, append-map, argmax, argmin, arithmetic-shift, arity-at-least-value, arity-at-least?, arity-checking-wrapper, arity-includes?, arity=?, arrow-contract-info-accepts-arglist, arrow-contract-info-chaperone-procedure, arrow-contract-info-check-first-order, arrow-contract-info?, asin, assert-unreachable, assf, assoc, assq, assv, assw, atan, banner, base->-doms/c, base->-rngs/c, base->?, bitwise-and, bitwise-bit-field, bitwise-bit-set?, bitwise-ior, bitwise-not, bitwise-xor, blame-add-car-context, blame-add-cdr-context, blame-add-missing-party, blame-add-nth-arg-context, blame-add-range-context, blame-add-unknown-context, blame-context, blame-contract, blame-fmt->-string, blame-missing-party?, blame-negative, blame-original?, blame-positive, blame-replace-negative, blame-replaced-negative?, blame-source, blame-swap, blame-swapped?, blame-update, blame-value, blame?, block-device-type-bits, boolean=?, boolean?, bound-identifier=?, box, box-cas!, box-immutable, box?, break-enabled, break-parameterization?, break-thread, build-chaperone-contract-property, build-compound-type-name, build-contract-property, build-flat-contract-property, build-list, build-path, build-path/convention-type, build-string, build-vector, byte-pregexp, byte-pregexp?, byte-ready?, byte-regexp, byte-regexp?, byte?, bytes, bytes->immutable-bytes, bytes->list, bytes->path, bytes->path-element, bytes->string/latin-1, bytes->string/locale, bytes->string/utf-8, bytes-append, bytes-append*, bytes-close-converter, bytes-convert, bytes-convert-end, bytes-converter?, bytes-copy, bytes-copy!, bytes-environment-variable-name?, bytes-fill!, bytes-join, bytes-length, bytes-no-nuls?, bytes-open-converter, bytes-ref, bytes-set!, bytes-utf-8-index, bytes-utf-8-length, bytes-utf-8-ref, bytes<?, bytes=?, bytes>?, bytes?, caaaar, caaadr, caaar, caadar, caaddr, caadr, caar, cadaar, cadadr, cadar, caddar, cadddr, caddr, cadr, call-in-continuation, call-in-nested-thread, call-with-break-parameterization, call-with-composable-continuation, call-with-continuation-barrier, call-with-continuation-prompt, call-with-current-continuation, call-with-default-reading-parameterization, call-with-escape-continuation, call-with-exception-handler, call-with-immediate-continuation-mark, call-with-input-bytes, call-with-input-string, call-with-output-bytes, call-with-output-string, call-with-parameterization, call-with-semaphore, call-with-semaphore/enable-break, call-with-values, call/cc, call/ec, car, cartesian-product, cdaaar, cdaadr, cdaar, cdadar, cdaddr, cdadr, cdar, cddaar, cddadr, cddar, cdddar, cddddr, cdddr, cddr, cdr, ceiling, channel-get, channel-put, channel-put-evt, channel-put-evt?, channel-try-get, channel?, chaperone-box, chaperone-channel, chaperone-continuation-mark-key, chaperone-contract-property?, chaperone-contract?, chaperone-evt, chaperone-hash, chaperone-hash-set, chaperone-of?, chaperone-procedure, chaperone-procedure*, chaperone-prompt-tag, chaperone-struct, chaperone-struct-type, chaperone-vector, chaperone-vector*, chaperone?, char->integer, char-alphabetic?, char-blank?, char-ci<=?, char-ci<?, char-ci=?, char-ci>=?, char-ci>?, char-downcase, char-extended-pictographic?, char-foldcase, char-general-category, char-grapheme-break-property, char-grapheme-step, char-graphic?, char-in, char-iso-control?, char-lower-case?, char-numeric?, char-punctuation?, char-ready?, char-symbolic?, char-title-case?, char-titlecase, char-upcase, char-upper-case?, char-utf-8-length, char-whitespace?, char<=?, char<?, char=?, char>=?, char>?, char?, character-device-type-bits, check-duplicate-identifier, checked-procedure-check-and-extract, choice-evt, class->interface, class-info, class-seal, class-unseal, class?, cleanse-path, close-input-port, close-output-port, coerce-chaperone-contract, coerce-chaperone-contracts, coerce-contract, coerce-contract/f, coerce-contracts, coerce-flat-contract, coerce-flat-contracts, collect-garbage, collection-file-path, collection-path, combinations, combine-output, compile, compile-allow-set!-undefined, compile-context-preservation-enabled, compile-enforce-module-constants, compile-syntax, compile-target-machine?, compiled-expression-add-target-machine, compiled-expression-recompile, compiled-expression?, compiled-module-expression?, complete-path?, complex/c, complex?, compose, compose1, conjoin, conjugate, cons, cons?, const, const*, continuation-mark-key?, continuation-mark-set->context, continuation-mark-set->iterator, continuation-mark-set->list, continuation-mark-set->list*, continuation-mark-set-first, continuation-mark-set?, continuation-marks, continuation-prompt-available?, continuation-prompt-tag?, continuation?, contract-continuation-mark-key, contract-custom-write-property-proc, contract-equivalent?, contract-first-order, contract-first-order-passes?, contract-late-neg-projection, contract-name, contract-proc, contract-projection, contract-property?, contract-random-generate, contract-random-generate-env?, contract-random-generate-fail, contract-random-generate-fail?, contract-random-generate-get-current-environment, contract-random-generate-stash, contract-random-generate/choose, contract-stronger?, contract-struct-exercise, contract-struct-generate, contract-struct-late-neg-projection, contract-struct-list-contract?, contract-val-first-projection, contract?, convert-stream, copy-file, copy-port, cos, cosh, count, current-blame-format, current-break-parameterization, current-code-inspector, current-command-line-arguments, current-compile, current-compile-realm, current-compile-target-machine, current-compiled-file-roots, current-continuation-marks, current-custodian, current-directory, current-directory-for-user, current-drive, current-environment-variables, current-error-message-adjuster, current-error-port, current-eval, current-evt-pseudo-random-generator, current-force-delete-permissions, current-future, current-gc-milliseconds, current-get-interaction-evt, current-get-interaction-input-port, current-inexact-milliseconds, current-inexact-monotonic-milliseconds, current-input-port, current-inspector, current-library-collection-links, current-library-collection-paths, current-load, current-load-extension, current-load-relative-directory, current-load/use-compiled, current-locale, current-logger, current-memory-use, current-milliseconds, current-module-declare-name, current-module-declare-source, current-module-name-resolver, current-module-path-for-load, current-namespace, current-output-port, current-parameterization, current-plumber, current-preserved-thread-cell-values, current-print, current-process-milliseconds, current-prompt-read, current-pseudo-random-generator, current-read-interaction, current-reader-guard, current-readtable, current-seconds, current-security-guard, current-subprocess-custodian-mode, current-subprocess-keep-file-descriptors, current-thread, current-thread-group, current-thread-initial-stack-size, current-write-relative-directory, curry, curryr, custodian-box-value, custodian-box?, custodian-limit-memory, custodian-managed-list, custodian-memory-accounting-available?, custodian-require-memory, custodian-shut-down?, custodian-shutdown-all, custodian?, custom-print-quotable-accessor, custom-print-quotable?, custom-write-accessor, custom-write-property-proc, custom-write?, date*-nanosecond, date*-time-zone-name, date*?, date-day, date-dst?, date-hour, date-minute, date-month, date-second, date-time-zone-offset, date-week-day, date-year, date-year-day, date?, datum->syntax, datum-intern-literal, default-continuation-prompt-tag, default-global-port-print-handler, degrees->radians, delete-directory, delete-file, denominator, dict-iter-contract, dict-key-contract, dict-value-contract, directory-exists?, directory-list, directory-type-bits, disjoin, display, displayln, double-flonum?, drop, drop-common-prefix, drop-right, dropf, dropf-right, dump-memory-stats, dup-input-port, dup-output-port, dynamic-get-field, dynamic-object/c, dynamic-require, dynamic-require-for-syntax, dynamic-send, dynamic-set-field!, dynamic-wind, eighth, empty, empty-sequence, empty-stream, empty?, environment-variables-copy, environment-variables-names, environment-variables-ref, environment-variables-set!, environment-variables?, eof, eof-object?, ephemeron-value, ephemeron?, eprintf, eq-contract-val, eq-contract?, eq-hash-code, eq?, equal-always-hash-code, equal-always-hash-code/recur, equal-always-secondary-hash-code, equal-always?, equal-always?/recur, equal-contract-val, equal-contract?, equal-hash-code, equal-hash-code/recur, equal-secondary-hash-code, equal<\%>, equal?, equal?/recur, eqv-hash-code, eqv?, error, error-contract->adjusted-string, error-display-handler, error-escape-handler, error-message->adjusted-string, error-message-adjuster-key, error-print-context-length, error-print-source-location, error-print-width, error-syntax->string-handler, error-value->string-handler, eval, eval-jit-enabled, eval-syntax, even?, evt/c, evt?, exact->inexact, exact-ceiling, exact-floor, exact-integer?, exact-nonnegative-integer?, exact-positive-integer?, exact-round, exact-truncate, exact?, executable-yield-handler, exit, exit-handler, exn-continuation-marks, exn-message, exn:break-continuation, exn:break:hang-up?, exn:break:terminate?, exn:break?, exn:fail:contract:arity?, exn:fail:contract:blame-object, exn:fail:contract:blame?, exn:fail:contract:continuation?, exn:fail:contract:divide-by-zero?, exn:fail:contract:non-fixnum-result?, exn:fail:contract:variable-id, exn:fail:contract:variable?, exn:fail:contract?, exn:fail:filesystem:errno-errno, exn:fail:filesystem:errno?, exn:fail:filesystem:exists?, exn:fail:filesystem:missing-module-path, exn:fail:filesystem:missing-module?, exn:fail:filesystem:version?, exn:fail:filesystem?, exn:fail:network:errno-errno, exn:fail:network:errno?, exn:fail:network?, exn:fail:object?, exn:fail:out-of-memory?, exn:fail:read-srclocs, exn:fail:read:eof?, exn:fail:read:non-char?, exn:fail:read?, exn:fail:syntax-exprs, exn:fail:syntax:missing-module-path, exn:fail:syntax:missing-module?, exn:fail:syntax:unbound?, exn:fail:syntax?, exn:fail:unsupported?, exn:fail:user?, exn:fail?, exn:misc:match?, exn:missing-module-accessor, exn:missing-module?, exn:srclocs-accessor, exn:srclocs?, exn?, exp, expand, expand-once, expand-syntax, expand-syntax-once, expand-syntax-to-top-form, expand-to-top-form, expand-user-path, explode-path, expt, externalizable<\%>, failure-result/c, false, false/c, false?, field-names, fifo-type-bits, fifth, file-exists?, file-name-from-path, file-or-directory-identity, file-or-directory-modify-seconds, file-or-directory-permissions, file-or-directory-stat, file-or-directory-type, file-position, file-position*, file-size, file-stream-buffer-mode, file-stream-port?, file-truncate, file-type-bits, filename-extension, filesystem-change-evt, filesystem-change-evt-cancel, filesystem-change-evt?, filesystem-root-list, filter, filter-map, filter-not, filter-read-input-port, find-compiled-file-roots, find-executable-path, find-library-collection-links, find-library-collection-paths, find-system-path, findf, first, fixnum?, flat-contract, flat-contract-predicate, flat-contract-property?, flat-contract?, flat-named-contract, flatten, floating-point-bytes->real, flonum?, floor, flush-output, fold-files, foldl, foldr, for-each, force, format, fourth, fprintf, free-identifier=?, free-label-identifier=?, free-template-identifier=?, free-transformer-identifier=?, fsemaphore-count, fsemaphore-post, fsemaphore-try-wait?, fsemaphore-wait, fsemaphore?, future, future?, futures-enabled?, gcd, generate-member-key, generate-temporaries, generic-set?, generic?, gensym, get-output-bytes, get-output-string, get/build-late-neg-projection, get/build-val-first-projection, getenv, global-port-print-handler, group-by, group-execute-bit, group-permission-bits, group-read-bit, group-write-bit, guard-evt, handle-evt, handle-evt?, has-blame?, has-contract?, hash, hash->list, hash-clear, hash-clear!, hash-copy, hash-count, hash-empty?, hash-ephemeron?, hash-eq?, hash-equal-always?, hash-equal?, hash-eqv?, hash-for-each, hash-has-key?, hash-iterate-first, hash-iterate-key, hash-iterate-key+value, hash-iterate-next, hash-iterate-pair, hash-iterate-value, hash-keys, hash-keys-subset?, hash-map, hash-placeholder?, hash-ref, hash-ref!, hash-ref-key, hash-remove, hash-remove!, hash-set, hash-set!, hash-set*, hash-set*!, hash-strong?, hash-update, hash-update!, hash-values, hash-weak?, hash?, hashalw, hasheq, hasheqv, identifier-binding, identifier-binding-portal-syntax, identifier-binding-symbol, identifier-distinct-binding, identifier-label-binding, identifier-prune-lexical-context, identifier-prune-to-source-module, identifier-remove-from-definition-context, identifier-template-binding, identifier-transformer-binding, identifier?, identity, if/c, imag-part, immutable?, impersonate-box, impersonate-channel, impersonate-continuation-mark-key, impersonate-hash, impersonate-hash-set, impersonate-procedure, impersonate-procedure*, impersonate-prompt-tag, impersonate-struct, impersonate-vector, impersonate-vector*, impersonator-contract?, impersonator-ephemeron, impersonator-of?, impersonator-prop:application-mark, impersonator-prop:blame, impersonator-prop:contracted, impersonator-property-accessor-procedure?, impersonator-property?, impersonator?, implementation?, implementation?/c, in-combinations, in-cycle, in-dict-pairs, in-parallel, in-permutations, in-sequences, in-values*-sequence, in-values-sequence, index-of, index-where, indexes-of, indexes-where, inexact->exact, inexact-real?, inexact?, infinite?, input-port?, inspector-superior?, inspector?, integer->char, integer->integer-bytes, integer-bytes->integer, integer-length, integer-sqrt, integer-sqrt/remainder, integer?, interface->method-names, interface-extension?, interface?, internal-definition-context-add-scopes, internal-definition-context-binding-identifiers, internal-definition-context-introduce, internal-definition-context-seal, internal-definition-context-splice-binding-identifier, internal-definition-context?, is-a?, is-a?/c, keyword->string, keyword-apply, keyword-apply/dict, keyword<?, keyword?, keywords-match, kill-thread, last, last-pair, lcm, length, liberal-define-context?, link-exists?, list, list*, list->bytes, list->mutable-set, list->mutable-setalw, list->mutable-seteq, list->mutable-seteqv, list->set, list->setalw, list->seteq, list->seteqv, list->string, list->vector, list->weak-set, list->weak-setalw, list->weak-seteq, list->weak-seteqv, list-contract?, list-prefix?, list-ref, list-set, list-tail, list-update, list?, listen-port-number?, load, load-extension, load-on-demand-enabled, load-relative, load-relative-extension, load/cd, load/use-compiled, local-expand, local-expand/capture-lifts, local-transformer-expand, local-transformer-expand/capture-lifts, locale-string-encoding, log, log-all-levels, log-level-evt, log-level?, log-max-level, log-message, log-receiver?, logger-name, logger?, magnitude, make-arity-at-least, make-base-empty-namespace, make-base-namespace, make-bytes, make-channel, make-chaperone-contract, make-continuation-mark-key, make-continuation-prompt-tag, make-contract, make-custodian, make-custodian-box, make-date, make-date*, make-derived-parameter, make-directory, make-directory*, make-do-sequence, make-empty-namespace, make-environment-variables, make-ephemeron, make-ephemeron-hash, make-ephemeron-hashalw, make-ephemeron-hasheq, make-ephemeron-hasheqv, make-exn, make-exn:break, make-exn:break:hang-up, make-exn:break:terminate, make-exn:fail, make-exn:fail:contract, make-exn:fail:contract:arity, make-exn:fail:contract:blame, make-exn:fail:contract:continuation, make-exn:fail:contract:divide-by-zero, make-exn:fail:contract:non-fixnum-result, make-exn:fail:contract:variable, make-exn:fail:filesystem, make-exn:fail:filesystem:errno, make-exn:fail:filesystem:exists, make-exn:fail:filesystem:missing-module, make-exn:fail:filesystem:version, make-exn:fail:network, make-exn:fail:network:errno, make-exn:fail:object, make-exn:fail:out-of-memory, make-exn:fail:read, make-exn:fail:read:eof, make-exn:fail:read:non-char, make-exn:fail:syntax, make-exn:fail:syntax:missing-module, make-exn:fail:syntax:unbound, make-exn:fail:unsupported, make-exn:fail:user, make-file-or-directory-link, make-flat-contract, make-fsemaphore, make-generic, make-hash, make-hash-placeholder, make-hashalw, make-hashalw-placeholder, make-hasheq, make-hasheq-placeholder, make-hasheqv, make-hasheqv-placeholder, make-immutable-hash, make-immutable-hashalw, make-immutable-hasheq, make-immutable-hasheqv, make-impersonator-property, make-input-port, make-input-port/read-to-peek, make-inspector, make-interned-syntax-introducer, make-keyword-procedure, make-known-char-range-list, make-list, make-lock-file-name, make-log-receiver, make-logger, make-mixin-contract, make-none/c, make-output-port, make-parameter, make-parent-directory*, make-phantom-bytes, make-pipe, make-pipe-with-specials, make-placeholder, make-plumber, make-polar, make-portal-syntax, make-prefab-struct, make-primitive-class, make-proj-contract, make-pseudo-random-generator, make-reader-graph, make-readtable, make-rectangular, make-rename-transformer, make-resolved-module-path, make-security-guard, make-semaphore, make-set!-transformer, make-shared-bytes, make-sibling-inspector, make-special-comment, make-srcloc, make-string, make-struct-field-accessor, make-struct-field-mutator, make-struct-type, make-struct-type-property, make-syntax-delta-introducer, make-syntax-introducer, make-tentative-pretty-print-output-port, make-thread-cell, make-thread-group, make-vector, make-weak-box, make-weak-hash, make-weak-hashalw, make-weak-hasheq, make-weak-hasheqv, make-will-executor, map, match-equality-test, matches-arity-exactly?, max, mcar, mcdr, mcons, member, member-name-key-hash-code, member-name-key=?, member-name-key?, memf, memory-order-acquire, memory-order-release, memq, memv, memw, merge-input, method-in-interface?, min, mixin-contract, module->exports, module->imports, module->indirect-exports, module->language-info, module->namespace, module->realm, module-cache-clear!, module-compiled-cross-phase-persistent?, module-compiled-exports, module-compiled-imports, module-compiled-indirect-exports, module-compiled-language-info, module-compiled-name, module-compiled-realm, module-compiled-submodules, module-declared?, module-path-index-join, module-path-index-resolve, module-path-index-split, module-path-index-submodule, module-path-index?, module-path?, module-predefined?, module-provide-protected?, modulo, mpair?, mutable-set, mutable-setalw, mutable-seteq, mutable-seteqv, n->th, nack-guard-evt, namespace-anchor->empty-namespace, namespace-anchor->namespace, namespace-anchor?, namespace-attach-module, namespace-attach-module-declaration, namespace-base-phase, namespace-call-with-registry-lock, namespace-mapped-symbols, namespace-module-identifier, namespace-module-registry, namespace-require, namespace-require/constant, namespace-require/copy, namespace-require/expansion-time, namespace-set-variable-value!, namespace-symbol->identifier, namespace-syntax-introduce, namespace-undefine-variable!, namespace-unprotect-module, namespace-variable-value, namespace?, nan?, natural-number/c, natural?, negate, negative-integer?, negative?, never-evt, newline, ninth, non-empty-string?, nonnegative-integer?, nonpositive-integer?, normal-case-path, normalize-arity, normalize-path, normalized-arity?, not, null, null?, number->string, number?, numerator, object\%, object->vector, object-info, object-interface, object-method-arity-includes?, object-name, object-or-false=?, object=-hash-code, object=?, object?, odd?, open-input-bytes, open-input-string, open-output-bytes, open-output-nowhere, open-output-string, order-of-magnitude, ormap, other-execute-bit, other-permission-bits, other-read-bit, other-write-bit, output-port?, pair?, parameter-procedure=?, parameter?, parameterization?, parse-command-line, partition, path->bytes, path->complete-path, path->directory-path, path->string, path-add-extension, path-add-suffix, path-convention-type, path-element->bytes, path-element->string, path-element?, path-for-some-system?, path-get-extension, path-has-extension?, path-list-string->path-list, path-only, path-replace-extension, path-replace-suffix, path-string?, path<?, path?, peek-byte, peek-byte-or-special, peek-bytes, peek-bytes!, peek-bytes-avail!, peek-bytes-avail!*, peek-bytes-avail!/enable-break, peek-char, peek-char-or-special, peek-string, peek-string!, permutations, phantom-bytes?, pi, pi.f, pipe-content-length, place-break, place-channel, place-channel-get, place-channel-put, place-channel-put/get, place-channel?, place-dead-evt, place-enabled?, place-kill, place-location?, place-message-allowed?, place-wait, place?, placeholder-get, placeholder-set!, placeholder?, plumber-add-flush!, plumber-flush-all, plumber-flush-handle-remove!, plumber-flush-handle?, plumber?, poll-guard-evt, port->list, port-closed-evt, port-closed?, port-commit-peeked, port-count-lines!, port-count-lines-enabled, port-counts-lines?, port-display-handler, port-file-identity, port-file-unlock, port-next-location, port-number?, port-print-handler, port-progress-evt, port-provides-progress-evts?, port-read-handler, port-try-file-lock?, port-waiting-peer?, port-write-handler, port-writes-atomic?, port-writes-special?, port?, portal-syntax-content, portal-syntax?, positive-integer?, positive?, predicate/c, prefab-key->struct-type, prefab-key?, prefab-struct-key, prefab-struct-type-key+field-count, preferences-lock-file-mode, pregexp, pregexp-quote, pregexp?, pretty-display, pretty-print, pretty-print-.-symbol-without-bars, pretty-print-abbreviate-read-macros, pretty-print-columns, pretty-print-current-style-table, pretty-print-depth, pretty-print-exact-as-decimal, pretty-print-extend-style-table, pretty-print-handler, pretty-print-newline, pretty-print-post-print-hook, pretty-print-pre-print-hook, pretty-print-print-hook, pretty-print-print-line, pretty-print-remap-stylable, pretty-print-show-inexactness, pretty-print-size-hook, pretty-print-style-table?, pretty-printing, pretty-write, primitive-closure?, primitive-result-arity, primitive?, print, print-as-expression, print-boolean-long-form, print-box, print-graph, print-hash-table, print-mpair-curly-braces, print-pair-curly-braces, print-reader-abbreviations, print-struct, print-syntax-width, print-unreadable, print-value-columns, print-vector-length, printable/c, printable<\%>, printf, println, procedure->method, procedure-arity, procedure-arity-includes?, procedure-arity-mask, procedure-arity?, procedure-closure-contents-eq?, procedure-extract-target, procedure-impersonator*?, procedure-keywords, procedure-realm, procedure-reduce-arity, procedure-reduce-arity-mask, procedure-reduce-keyword-arity, procedure-reduce-keyword-arity-mask, procedure-rename, procedure-result-arity, procedure-specialize, procedure-struct-type?, procedure?, processor-count, progress-evt?, promise-forced?, promise-running?, promise/name?, promise?, prop:arity-string, prop:arrow-contract, prop:arrow-contract-get-info, prop:arrow-contract?, prop:authentic, prop:blame, prop:chaperone-contract, prop:checked-procedure, prop:contract, prop:contracted, prop:custom-print-quotable, prop:custom-write, prop:dict, prop:equal+hash, prop:evt, prop:exn:missing-module, prop:exn:srclocs, prop:expansion-contexts, prop:flat-contract, prop:impersonator-of, prop:input-port, prop:liberal-define-context, prop:object-name, prop:orc-contract, prop:orc-contract-get-subcontracts, prop:orc-contract?, prop:output-port, prop:place-location, prop:procedure, prop:recursive-contract, prop:recursive-contract-unroll, prop:recursive-contract?, prop:rename-transformer, prop:sealed, prop:sequence, prop:set!-transformer, prop:stream, proper-subset?, pseudo-random-generator->vector, pseudo-random-generator-vector?, pseudo-random-generator?, put-preferences, putenv, quotient, quotient/remainder, radians->degrees, raise, raise-argument-error, raise-argument-error*, raise-arguments-error, raise-arguments-error*, raise-arity-error, raise-arity-error*, raise-arity-mask-error, raise-arity-mask-error*, raise-contract-error, raise-mismatch-error, raise-range-error, raise-range-error*, raise-result-arity-error, raise-result-arity-error*, raise-result-error, raise-result-error*, raise-type-error, raise-user-error, random, random-seed, rational?, rationalize, read, read-accept-bar-quote, read-accept-box, read-accept-compiled, read-accept-dot, read-accept-graph, read-accept-infix-dot, read-accept-lang, read-accept-quasiquote, read-accept-reader, read-byte, read-byte-or-special, read-bytes, read-bytes!, read-bytes-avail!, read-bytes-avail!*, read-bytes-avail!/enable-break, read-bytes-line, read-case-sensitive, read-cdot, read-char, read-char-or-special, read-curly-brace-as-paren, read-curly-brace-with-tag, read-decimal-as-inexact, read-eval-print-loop, read-installation-configuration-table, read-language, read-line, read-on-demand-source, read-single-flonum, read-square-bracket-as-paren, read-square-bracket-with-tag, read-string, read-string!, read-syntax, read-syntax-accept-graph, read-syntax/recursive, read/recursive, readtable-mapping, readtable?, real->decimal-string, real->double-flonum, real->floating-point-bytes, real->single-flonum, real-part, real?, reencode-input-port, reencode-output-port, regexp, regexp-match, regexp-match-exact?, regexp-match-peek, regexp-match-peek-immediate, regexp-match-peek-positions, regexp-match-peek-positions-immediate, regexp-match-peek-positions-immediate/end, regexp-match-peek-positions/end, regexp-match-positions, regexp-match-positions/end, regexp-match/end, regexp-match?, regexp-max-lookbehind, regexp-quote, regexp-replace, regexp-replace*, regexp-replace-quote, regexp-replaces, regexp-split, regexp-try-match, regexp?, regular-file-type-bits, relative-path?, remainder, remf, remf*, remove, remove*, remq, remq*, remv, remv*, remw, remw*, rename-contract, rename-file-or-directory, rename-transformer-target, rename-transformer?, replace-evt, reroot-path, resolve-path, resolved-module-path-name, resolved-module-path?, rest, reverse, round, second, seconds->date, security-guard?, semaphore-peek-evt, semaphore-peek-evt?, semaphore-post, semaphore-try-wait?, semaphore-wait, semaphore-wait/enable-break, semaphore?, sequence->list, sequence->stream, sequence-add-between, sequence-andmap, sequence-append, sequence-count, sequence-filter, sequence-fold, sequence-for-each, sequence-generate, sequence-generate*, sequence-length, sequence-map, sequence-ormap, sequence-ref, sequence-tail, sequence/c, sequence?, set, set!-transformer-procedure, set!-transformer?, set->list, set->stream, set-add, set-add!, set-box!, set-box*!, set-clear, set-clear!, set-copy, set-copy-clear, set-count, set-empty?, set-eq?, set-equal-always?, set-equal?, set-eqv?, set-first, set-for-each, set-group-id-bit, set-implements/c, set-implements?, set-intersect, set-intersect!, set-map, set-mcar!, set-mcdr!, set-member?, set-mutable?, set-phantom-bytes!, set-port-next-location!, set-remove, set-remove!, set-rest, set-subtract, set-subtract!, set-symmetric-difference, set-symmetric-difference!, set-union, set-union!, set-user-id-bit, set-weak?, set=?, set?, setalw, seteq, seteqv, seventh, sgn, sha1-bytes, sha224-bytes, sha256-bytes, shared-bytes, shell-execute, shrink-path-wrt, shuffle, simple-form-path, simplify-path, sin, single-flonum-available?, single-flonum?, sinh, sixth, skip-projection-wrapper?, sleep, socket-type-bits, some-system-path->string, special-comment-value, special-comment?, special-filter-input-port, split-at, split-at-right, split-common-prefix, split-path, splitf-at, splitf-at-right, sqr, sqrt, srcloc->string, srcloc-column, srcloc-line, srcloc-position, srcloc-source, srcloc-span, srcloc?, stencil-vector, stencil-vector-length, stencil-vector-mask, stencil-vector-mask-width, stencil-vector-ref, stencil-vector-set!, stencil-vector-update, stencil-vector?, sticky-bit, stop-after, stop-before, stream->list, stream-add-between, stream-andmap, stream-append, stream-count, stream-empty?, stream-filter, stream-first, stream-fold, stream-for-each, stream-force, stream-length, stream-map, stream-ormap, stream-ref, stream-rest, stream-tail, stream-take, stream/c, stream?, string, string->bytes/latin-1, string->bytes/locale, string->bytes/utf-8, string->immutable-string, string->keyword, string->list, string->number, string->path, string->path-element, string->some-system-path, string->symbol, string->uninterned-symbol, string->unreadable-symbol, string-append, string-append*, string-append-immutable, string-ci<=?, string-ci<?, string-ci=?, string-ci>=?, string-ci>?, string-contains?, string-copy, string-copy!, string-downcase, string-environment-variable-name?, string-fill!, string-foldcase, string-grapheme-count, string-grapheme-span, string-length, string-locale-ci<?, string-locale-ci=?, string-locale-ci>?, string-locale-downcase, string-locale-upcase, string-locale<?, string-locale=?, string-locale>?, string-no-nuls?, string-normalize-nfc, string-normalize-nfd, string-normalize-nfkc, string-normalize-nfkd, string-port?, string-prefix?, string-ref, string-set!, string-suffix?, string-titlecase, string-upcase, string-utf-8-length, string<=?, string<?, string=?, string>=?, string>?, string?, struct->vector, struct-accessor-procedure?, struct-constructor-procedure?, struct-info, struct-mutator-procedure?, struct-predicate-procedure?, struct-type-authentic?, struct-type-info, struct-type-make-constructor, struct-type-make-predicate, struct-type-property-accessor-procedure?, struct-type-property-predicate-procedure?, struct-type-property/c, struct-type-property?, struct-type-sealed?, struct-type?, struct:arity-at-least, struct:arrow-contract-info, struct:date, struct:date*, struct:exn, struct:exn:break, struct:exn:break:hang-up, struct:exn:break:terminate, struct:exn:fail, struct:exn:fail:contract, struct:exn:fail:contract:arity, struct:exn:fail:contract:blame, struct:exn:fail:contract:continuation, struct:exn:fail:contract:divide-by-zero, struct:exn:fail:contract:non-fixnum-result, struct:exn:fail:contract:variable, struct:exn:fail:filesystem, struct:exn:fail:filesystem:errno, struct:exn:fail:filesystem:exists, struct:exn:fail:filesystem:missing-module, struct:exn:fail:filesystem:version, struct:exn:fail:network, struct:exn:fail:network:errno, struct:exn:fail:object, struct:exn:fail:out-of-memory, struct:exn:fail:read, struct:exn:fail:read:eof, struct:exn:fail:read:non-char, struct:exn:fail:syntax, struct:exn:fail:syntax:missing-module, struct:exn:fail:syntax:unbound, struct:exn:fail:unsupported, struct:exn:fail:user, struct:srcloc, struct?, sub1, subbytes, subclass?, subclass?/c, subprocess, subprocess-group-enabled, subprocess-kill, subprocess-pid, subprocess-status, subprocess-wait, subprocess?, subset?, substring, suggest/c, symbol->string, symbol-interned?, symbol-unreadable?, symbol<?, symbol=?, symbol?, symbolic-link-type-bits, sync, sync/enable-break, sync/timeout, sync/timeout/enable-break, syntax->datum, syntax->list, syntax-arm, syntax-binding-set, syntax-binding-set->syntax, syntax-binding-set?, syntax-bound-phases, syntax-bound-symbols, syntax-column, syntax-debug-info, syntax-disarm, syntax-e, syntax-line, syntax-local-apply-transformer, syntax-local-bind-syntaxes, syntax-local-certifier, syntax-local-context, syntax-local-expand-expression, syntax-local-get-shadower, syntax-local-identifier-as-binding, syntax-local-introduce, syntax-local-lift-context, syntax-local-lift-expression, syntax-local-lift-module, syntax-local-lift-module-end-declaration, syntax-local-lift-provide, syntax-local-lift-require, syntax-local-lift-values-expression, syntax-local-make-definition-context, syntax-local-make-definition-context-introducer, syntax-local-make-delta-introducer, syntax-local-module-defined-identifiers, syntax-local-module-exports, syntax-local-module-interned-scope-symbols, syntax-local-module-required-identifiers, syntax-local-name, syntax-local-phase-level, syntax-local-submodules, syntax-local-transforming-module-provides?, syntax-local-value, syntax-local-value/immediate, syntax-original?, syntax-position, syntax-property, syntax-property-preserved?, syntax-property-remove, syntax-property-symbol-keys, syntax-protect, syntax-rearm, syntax-recertify, syntax-shift-phase-level, syntax-source, syntax-source-module, syntax-span, syntax-taint, syntax-tainted?, syntax-track-origin, syntax-transforming-module-expression?, syntax-transforming-with-lifts?, syntax-transforming?, syntax?, system-big-endian?, system-idle-evt, system-language+country, system-library-subpath, system-path-convention-type, system-type, tail-marks-match?, take, take-common-prefix, take-right, takef, takef-right, tan, tanh, tcp-abandon-port, tcp-accept, tcp-accept-evt, tcp-accept-ready?, tcp-accept/enable-break, tcp-addresses, tcp-close, tcp-connect, tcp-connect/enable-break, tcp-listen, tcp-listener?, tcp-port?, tentative-pretty-print-port-cancel, tentative-pretty-print-port-transfer, tenth, terminal-port?, the-unsupplied-arg, third, thread, thread-cell-ref, thread-cell-set!, thread-cell-values?, thread-cell?, thread-dead-evt, thread-dead?, thread-group?, thread-receive, thread-receive-evt, thread-resume, thread-resume-evt, thread-rewind-receive, thread-running?, thread-send, thread-suspend, thread-suspend-evt, thread-try-receive, thread-wait, thread/suspend-to-kill, thread?, time-apply, touch, true, truncate, udp-addresses, udp-bind!, udp-bound?, udp-close, udp-connect!, udp-connected?, udp-multicast-interface, udp-multicast-join-group!, udp-multicast-leave-group!, udp-multicast-loopback?, udp-multicast-set-interface!, udp-multicast-set-loopback!, udp-multicast-set-ttl!, udp-multicast-ttl, udp-open-socket, udp-receive!, udp-receive!*, udp-receive!-evt, udp-receive!/enable-break, udp-receive-ready-evt, udp-send, udp-send*, udp-send-evt, udp-send-ready-evt, udp-send-to, udp-send-to*, udp-send-to-evt, udp-send-to/enable-break, udp-send/enable-break, udp-set-receive-buffer-size!, udp-set-ttl!, udp-ttl, udp?, unbox, unbox*, uncaught-exception-handler, unit?, unquoted-printing-string, unquoted-printing-string-value, unquoted-printing-string?, unspecified-dom, unsupplied-arg?, use-collection-link-paths, use-compiled-file-check, use-compiled-file-paths, use-user-specific-search-paths, user-execute-bit, user-permission-bits, user-read-bit, user-write-bit, value-blame, value-contract, values, variable-reference->empty-namespace, variable-reference->module-base-phase, variable-reference->module-declaration-inspector, variable-reference->module-path-index, variable-reference->module-source, variable-reference->namespace, variable-reference->phase, variable-reference->resolved-module-path, variable-reference-constant?, variable-reference-from-unsafe?, variable-reference?, vector, vector*-append, vector*-copy, vector*-extend, vector*-length, vector*-ref, vector*-set!, vector*-set/copy, vector->immutable-vector, vector->list, vector->pseudo-random-generator, vector->pseudo-random-generator!, vector->values, vector-append, vector-argmax, vector-argmin, vector-cas!, vector-copy, vector-copy!, vector-count, vector-drop, vector-drop-right, vector-empty?, vector-extend, vector-fill!, vector-filter, vector-filter-not, vector-immutable, vector-length, vector-map, vector-map!, vector-member, vector-memq, vector-memv, vector-ref, vector-set!, vector-set*!, vector-set-performance-stats!, vector-set/copy, vector-split-at, vector-split-at-right, vector-take, vector-take-right, vector?, version, void, void?, weak-box-value, weak-box?, weak-set, weak-setalw, weak-seteq, weak-seteqv, will-execute, will-executor?, will-register, will-try-execute, with-input-from-bytes, with-input-from-string, with-output-to-bytes, with-output-to-string, would-be-future, wrap-evt, writable<\%>, write, write-byte, write-bytes, write-bytes-avail, write-bytes-avail*, write-bytes-avail-evt, write-bytes-avail/enable-break, write-char, write-special, write-special-avail*, write-special-evt, write-string, writeln, xor, zero?, check-tail-contract, for-clause-syntax-protect, legacy-match-expander?, match-...-nesting, match-expander?, prop:legacy-match-expander, prop:match-expander, syntax-local-match-introduce, syntax-pattern-variable?, \#\%app, \#\%datum, \#\%declare, \#\%expression, \#\%module-begin, \#\%plain-app, \#\%plain-lambda, \#\%plain-module-begin, \#\%printing-module-begin, \#\%provide, \#\%require, \#\%stratified-body, \#\%top, \#\%top-interaction, \#\%variable-reference, ->, ->*, ->*m, ->d, ->dm, ->i, ->m, ..., :do-in, <=/c, =/c, ==, =>, >=/c, _, absent, abstract, add-between, all-defined-out, all-from-out, and, and/c, any, any/c, apply, arity-at-least, arrow-contract-info, augment, augment*, augment-final, augment-final*, augride, augride*, bad-number-of-results, begin, begin-for-syntax, begin0, between/c, blame-add-context, box-immutable/c, box/c, call-with-atomic-output-file, call-with-file-lock/timeout, call-with-input-file, call-with-input-file*, call-with-output-file, call-with-output-file*, case, case->, case->m, case-lambda, channel/c, char-in/c, check-duplicates, class, class*, class-field-accessor, class-field-mutator, class/c, class/derived, combine-in, combine-out, command-line, compound-unit, compound-unit/infer, cond, cons/c, cons/dc, continuation-mark-key/c, contract, contract-exercise, contract-in, contract-out, contract-pos/neg-doubling, contract-struct, contracted, copy-directory/files, current-contract-region, date, date*, define, define-compound-unit, define-compound-unit/infer, define-contract-struct, define-custom-hash-types, define-custom-set-types, define-for-syntax, define-local-member-name, define-logger, define-match-expander, define-member-name, define-module-boundary-contract, define-namespace-anchor, define-opt/c, define-sequence-syntax, define-serializable-class, define-serializable-class*, define-signature, define-signature-form, define-splicing-for-clause-syntax, define-struct, define-struct/contract, define-struct/derived, define-syntax, define-syntax-rule, define-syntaxes, define-unit, define-unit-binding, define-unit-from-context, define-unit/contract, define-unit/new-import-export, define-unit/s, define-values, define-values-for-export, define-values-for-syntax, define-values/invoke-unit, define-values/invoke-unit/infer, define/augment, define/augment-final, define/augride, define/contract, define/final-prop, define/match, define/overment, define/override, define/override-final, define/private, define/public, define/public-final, define/pubment, define/subexpression-pos-prop, define/subexpression-pos-prop/name, delay, delay/idle, delay/name, delay/strict, delay/sync, delay/thread, delete-directory/files, dict->list, dict-can-functional-set?, dict-can-remove-keys?, dict-clear, dict-clear!, dict-copy, dict-count, dict-empty?, dict-for-each, dict-has-key?, dict-implements/c, dict-implements?, dict-iterate-first, dict-iterate-key, dict-iterate-next, dict-iterate-value, dict-keys, dict-map, dict-map/copy, dict-mutable?, dict-ref, dict-ref!, dict-remove, dict-remove!, dict-set, dict-set!, dict-set*, dict-set*!, dict-update, dict-update!, dict-values, dict?, display-lines, display-lines-to-file, display-to-file, do, dynamic->*, dynamic-instantiate, dynamic-place, dynamic-place*, else, eof-evt, except, except-in, except-out, exn, exn:break, exn:break:hang-up, exn:break:terminate, exn:fail, exn:fail:contract, exn:fail:contract:arity, exn:fail:contract:blame, exn:fail:contract:continuation, exn:fail:contract:divide-by-zero, exn:fail:contract:non-fixnum-result, exn:fail:contract:variable, exn:fail:filesystem, exn:fail:filesystem:errno, exn:fail:filesystem:exists, exn:fail:filesystem:missing-module, exn:fail:filesystem:version, exn:fail:network, exn:fail:network:errno, exn:fail:object, exn:fail:out-of-memory, exn:fail:read, exn:fail:read:eof, exn:fail:read:non-char, exn:fail:syntax, exn:fail:syntax:missing-module, exn:fail:syntax:unbound, exn:fail:unsupported, exn:fail:user, export, extends, failure-cont, field, field-bound?, file, file->bytes, file->bytes-lines, file->lines, file->list, file->string, file->value, find-files, find-relative-path, first-or/c, flat-contract-with-explanation, flat-murec-contract, flat-rec-contract, for, for*, for*/and, for*/async, for*/first, for*/fold, for*/fold/derived, for*/foldr, for*/foldr/derived, for*/hash, for*/hashalw, for*/hasheq, for*/hasheqv, for*/last, for*/list, for*/list/concurrent, for*/lists, for*/mutable-set, for*/mutable-setalw, for*/mutable-seteq, for*/mutable-seteqv, for*/or, for*/product, for*/set, for*/setalw, for*/seteq, for*/seteqv, for*/stream, for*/sum, for*/vector, for*/weak-set, for*/weak-setalw, for*/weak-seteq, for*/weak-seteqv, for-label, for-meta, for-space, for-syntax, for-template, for/and, for/async, for/first, for/fold, for/fold/derived, for/foldr, for/foldr/derived, for/hash, for/hashalw, for/hasheq, for/hasheqv, for/last, for/list, for/list/concurrent, for/lists, for/mutable-set, for/mutable-setalw, for/mutable-seteq, for/mutable-seteqv, for/or, for/product, for/set, for/setalw, for/seteq, for/seteqv, for/stream, for/sum, for/vector, for/weak-set, for/weak-setalw, for/weak-seteq, for/weak-seteqv, gen:custom-write, gen:dict, gen:equal+hash, gen:equal-mode+hash, gen:set, gen:stream, generic, get-field, get-preference, hash-copy-clear, hash-map/copy, hash/c, hash/dc, if, implies, import, in-bytes, in-bytes-lines, in-dict, in-dict-keys, in-dict-values, in-directory, in-ephemeron-hash, in-ephemeron-hash-keys, in-ephemeron-hash-pairs, in-ephemeron-hash-values, in-hash, in-hash-keys, in-hash-pairs, in-hash-values, in-immutable-hash, in-immutable-hash-keys, in-immutable-hash-pairs, in-immutable-hash-values, in-immutable-set, in-inclusive-range, in-indexed, in-input-port-bytes, in-input-port-chars, in-lines, in-list, in-mlist, in-mutable-hash, in-mutable-hash-keys, in-mutable-hash-pairs, in-mutable-hash-values, in-mutable-set, in-naturals, in-port, in-producer, in-range, in-set, in-slice, in-stream, in-string, in-syntax, in-value, in-vector, in-weak-hash, in-weak-hash-keys, in-weak-hash-pairs, in-weak-hash-values, in-weak-set, include, include-at/relative-to, include-at/relative-to/reader, include/reader, inclusive-range, inherit, inherit-field, inherit/inner, inherit/super, init, init-depend, init-field, init-rest, initiate-sequence, inner, input-port-append, inspect, instanceof/c, instantiate, integer-in, interface, interface*, invariant-assertion, invoke-unit, invoke-unit/infer, lambda, lazy, let, let*, let*-values, let-syntax, let-syntaxes, let-values, let/cc, let/ec, letrec, letrec-syntax, letrec-syntaxes, letrec-syntaxes+values, letrec-values, lib, link, list*of, list/c, listof, local, local-require, log-debug, log-error, log-fatal, log-info, log-warning, make-custom-hash, make-custom-hash-types, make-custom-set, make-custom-set-types, make-handle-get-preference-locked, make-immutable-custom-hash, make-limited-input-port, make-mutable-custom-set, make-object, make-temporary-directory, make-temporary-directory*, make-temporary-file, make-temporary-file*, make-weak-custom-hash, make-weak-custom-set, match, match*, match*/derived, match-define, match-define-values, match-lambda, match-lambda*, match-lambda**, match-let, match-let*, match-let*-values, match-let-values, match-letrec, match-letrec-values, match/derived, match/values, member-name-key, mixin, module, module*, module+, mutable-treelist/c, nand, new, new-∀/c, new-∃/c, non-empty-listof, none/c, nor, not/c, object-contract, object/c, one-of/c, only, only-in, only-meta-in, only-space-in, open, open-input-file, open-input-output-file, open-output-file, opt/c, or, or/c, overment, overment*, override, override*, override-final, override-final*, parameter/c, parameterize, parameterize*, parameterize-break, parametric->/c, pathlist-closure, peek-bytes!-evt, peek-bytes-avail!-evt, peek-bytes-evt, peek-string!-evt, peek-string-evt, peeking-input-port, place, place*, place/context, planet, port->bytes, port->bytes-lines, port->lines, port->string, prefix, prefix-in, prefix-out, pretty-format, private, private*, procedure-arity-includes/c, process, process*, process*/ports, process/ports, promise/c, prompt-tag/c, prop:dict/contract, property/c, protect-out, provide, provide-signature-elements, provide/contract, public, public*, public-final, public-final*, pubment, pubment*, quasiquote, quasisyntax, quasisyntax/loc, quote, quote-syntax, quote-syntax/prune, raise-blame-error, raise-not-cons-blame-error, raise-syntax-error, range, read-bytes!-evt, read-bytes-avail!-evt, read-bytes-evt, read-bytes-line-evt, read-line-evt, read-string!-evt, read-string-evt, real-in, recontract-out, recursive-contract, regexp-match*, regexp-match-evt, regexp-match-peek-positions*, regexp-match-positions*, relative-in, relocate-input-port, relocate-output-port, remove-duplicates, rename, rename-in, rename-inner, rename-out, rename-super, require, send, send*, send+, send-generic, send/apply, send/keyword-apply, set!, set!-values, set-field!, set/c, shared, sort, srcloc, stream, stream*, stream-cons, stream-lazy, string-join, string-len/c, string-normalize-spaces, string-replace, string-split, string-trim, struct, struct*, struct-copy, struct-field-index, struct-guard/c, struct-out, struct/c, struct/contract, struct/ctc, struct/dc, struct/derived, submod, super, super-instantiate, super-make-object, super-new, symbols, syntax, syntax-binding-set-extend, syntax-case, syntax-case*, syntax-deserialize, syntax-id-rules, syntax-rules, syntax-serialize, syntax/c, syntax/loc, system, system*, system*/exit-code, system/exit-code, tag, this, this\%, thunk, thunk*, time, transplant-input-port, transplant-output-port, treelist/c, unconstrained-domain->, unit, unit-from-context, unit/c, unit/new-import-export, unit/s, unless, unquote, unquote-splicing, unsyntax, unsyntax-splicing, values/drop, vector-immutable/c, vector-immutableof, vector-sort, vector-sort!, vector/c, vectorof, when, with-continuation-mark, with-contract, with-contract-continuation-mark, with-handlers, with-handlers*, with-input-from-file, with-method, with-output-to-file, with-syntax, write-to-file, ~.a, ~.s, ~.v, ~?, ~@, ~a, ~e, ~r, ~s, ~v, λ, equal?, equal, eq
		},
		keywordstyle=\color{blue},
		classoffset=3,
		morekeywords={import, export},
		keywordstyle=\color{green},
		classoffset=4,
		morekeywords={fsm, make-dfa, sm-apply, sm-states, sm-sigma, sm-start, sm-finals, sm-rules, sm-gamma, sm-showtransitions, sm-graph, gen-state, make-ndpda, make-ndfa, gen-regexp-word, singleton-regexp, kleenestar-regexp, concat-regexp, union-regexp, make-cfg, make-rg, make-csg, grammar-derive, grammar-not-derive, check-accept?, check-reject?},
		keywordstyle=\color{pakistangreen},
		classoffset=5,
		morekeywords={check-equal?, check-expect, check-equal, check-pred},
		keywordstyle=\color{sblue},
		classoffset=0,
		alsoletter={',`,-,/,>,<,\#,\%,?,=,*},
		moredelim=**[is][\color{lightgray}]{<<@<<}{>>@>>},
		moredelim=**[is][\itshape\color{OliveGreen}]{<<;<<}{>>;>>},
	}

\definecolor{sblue}{rgb}{0.14, 0.16, 0.48}
\definecolor{regalia}{rgb}{0.32, 0.18, 0.5}
\definecolor{palatinatepurple}{rgb}{0.41, 0.16, 0.38}
\definecolor{pakistangreen}{rgb}{0.0, 0.4, 0.0}
\definecolor{darkorange}{rgb}{1.0, 0.55, 0.0}

\newcommand{\fsm}{\texttt{FSM}}
\newcommand{\dfa}{\texttt{dfa}}
\newcommand{\ndfa}{\texttt{ndfa}}
\newcommand{\pda}{\texttt{pda}}

\newcommand{\sig}{\texttt{\(\Sigma\)}}
\newcommand{\gam}{\texttt{\(\Gamma\)}}

\newcommand{\delt}{\texttt{\(\delta\)}}

\newcommand{\quot}{\texttt{\textquotesingle{}}}

\newcommand{\step}{\texttt{$\vdash$}}

\newcommand{\accept}{\texttt{\textquotesingle{}accept}}
\newcommand{\reject}{\texttt{\textquotesingle{}reject}}

\newcommand{\arrow}{\(\rightarrow\)}
\newcommand{\flatt}{\texttt{FLAT}}

\newcommand{\gviz}{\texttt{GraphViz}}
\newcommand{\jflap}{\texttt{JFLAP}}
\newcommand{\lang}{\texttt{L = ab$^*$$ \ \cup \ $(ab)$^*$b$^*$}}

\definecolor{deepsaffron}{rgb}{1.0, 0.6, 0.2}
\definecolor{darkgreen}{RGB}{102,170,102}
\definecolor{ForestGreen}{RGB}{34,139,34}
\definecolor{amber}{rgb}{1.0, 0.75, 0.0}
\definecolor{deeppink}{rgb}{1.0, 0.08, 0.58}
\definecolor{desert}{rgb}{0.76, 0.6, 0.42}
\definecolor{dukeblue}{rgb}{0.0, 0.0, 0.61}
\definecolor{darkpastelgreen}{rgb}{0.01, 0.75, 0.24}
\definecolor{olivegreen}{RGB}{34, 139, 34}

\title{Visual Execution and Validation of Finite-State Machines and Pushdown Automata}

\author{Marco T. Moraz\'{a}n \and David Anthony K. Fields \and Andr\'{e}s M. Garced \and Tijana Mini\'{c}}

\author{Marco T. Moraz\'{a}n
\institute{Seton Hall University}
\email{morazanm@shu.edu}
\and
David Anthony K. Fields
\institute{Seton Hall University}
\email{fieldsda@shu.edu}
\and
Andr\'{e}s M. Garced
\institute{Seton Hall University}
\email{maldona2@shu.edu}
\and
Tijana Mini\'{c}
\institute{University of Washington}
\email{tminic@uw.edu}
}

\def\titlerunning{Visualizing Finite-State Machine and Pushdown Automata Execution}  
\def\authorrunning{Moraz\'{a}n, Fields, Garced, and Mini\'{c} }

\begin{document}
	
\maketitle

\begin{abstract}
In Formal Languages and Automata Theory courses, students find understanding nondeterministic finite-state and pushdown automata difficult. In many cases, this means that it is challenging for them to comprehend the operational semantics of such machines and, as a consequence, determine why a word is accepted or rejected. This is not entirely surprising, because students are mostly trained to design and implement deterministic programs. Comprehension of pushdown automata is further complicated, because reasoning about the stack is necessary. A common difficulty students face, for example, is understanding that two different computations on the same word may reach the same state with different stack values. To aid student understanding, we present two novel dynamic visualization tools for \fsm{}--a domain-specific programming language for the Automata Theory classroom--to support the design of such machines. These tools visualize all computations that may be performed, respectively, by a nondeterministic finite-state machine or by a pushdown automata in a stepwise manner. In addition, these tools aid the machine verification process by allowing users to visually validate whether the properties a state represents hold when a machine transitions into it.
\end{abstract}
	
\section{Introduction}

In Formal Languages and Automata Theory (\flatt{}) courses, one of the biggest challenges students face is understanding nondeterminism. This means students struggle to grasp why a given word is or is not in the language of a machine. Usually, students are forced to take this challenge head-on when they are introduced to nondeterministic finite-state automata (\ndfa{}) and again when they are exposed to pushdown automata (\pda{}). Such machines may have a choice when transitioning from their current state to the next state. This choice is only made if a transition may lead to an accepting configuration. Otherwise, none of the transitions are performed and the machine halts. Nondeterministic machines use their ``powerful intuition'' to decide which transition, if any, to perform. Not surprisingly, this is counterintuitive to students (mostly) trained in deterministic programming.

At first, few, if any, students can imagine how this ``powerful intuition'' may be simulated or implemented. Many textbooks explain this is possible, given a machine and a word, by performing a search on a computation tree either sequentially (i.e., a breadth-first search) or in parallel \cite{Gurari,Hopcroft,Lewis,Linz,Martin,PBFLAT,Rich,Sipser}. A computation tree has a path for every computation a machine may perform. If an accepting configuration is found during this search then the machine accepts. If all paths terminate without finding an accepting configuration, then the machine rejects. Otherwise, the machine never terminates. Multiple paths in a computation tree arise from nondeterministic choices the machine has at any given step. Initially, many \flatt{} students find this counterintuitive, because they assume nondeterministic machines have the same operational semantics as deterministic finite-state automata (\dfa{}). That is, they assume (despite having been told the contrary) that the machine's transitions rules alone must specify how the machine changes state. To further compound the comprehension problem, some students feel uncomfortable with machines that do not always consume the entire input word as is done by \dfa{}s.

A modern approach to make understanding the operational semantics of \ndfa{}s and \pda{}s easier is to have students implement them as programs using \fsm{}--a domain-specific language embedded in \texttt{Racket} \cite{Racket} for the \flatt{} classroom \cite{fsm}. This approach presents students with a design recipe for state machines that guides them through the process of designing, implementing, validating, and verifying state machines, including \ndfa{}s and \pda{}s. Design recipes, first developed by Felleisen et al. \cite{HtDP2} and later expanded by Moraz\'{a}n \cite{APS,APD} to instruct programming beginners, present a series of concrete steps a student needs to satisfy during the software development process. In addition, \fsm{} provides, for example, built-in visualization tools to generate the transition diagram of a given \ndfa{} or \pda{} and to visualize the dynamic application of a machine to a given word in the machine's language \cite{fsm-viz}.

Despite the scaffolding provided by \fsm{}, some students still struggle to understand how machines like \ndfa{}s and \pda{}s operate. The existing \fsm{} visualization tool only simulates a single accepting computation that may be performed by a nondeterministic machine. This helps students understand why a word is accepted. Not surprisingly, the tool is of no or little help in understanding why a word is rejected. To address this problem, we developed new dynamic visualization tools that display, respectively, the stepwise application of an \ndfa{} and a \pda{} to a given word (regardless of whether or not the word is in the machine's language). In essence, using a breadth-first traversal, the tools trace all paths in a computation tree that do not share a configuration. When an already visited configuration is encountered the computation it belongs to is no longer explored. This is safe to do because exploring a repeated configuration results in replicating a subtree that exists elsewhere in the computation tree. The tools are designed with the following goals in mind:
\begin{itemize}
  \item[$\circ$] Visualize computations using the given machine's transition diagram
  \item[$\circ$] Highlight the last transition used in all displayed computations
  \item[$\circ$] Move forward and backwards in the computations displayed
  \item[$\circ$] Display the number of computations without repeated configurations
  \item[$\circ$] Highlight, if any, accepting computations
  \item[$\circ$] Visually validate the design role of states using node coloring
  \item[$\circ$] Move in one step to the next/previous configuration for which a state design role does not hold
  \item[$\circ$] Present a low extraneous cognitive load
  \item[$\circ$] Optionally add a dead state to guarantee the consumption of the entire given word
\end{itemize}
In addition, our design is guided by the Norman principles for effective design \cite{Norman}. In our context, effective design means a user-friendly and easy-to-use interface. These principles include:
\begin{description}[leftmargin=!,labelwidth=\widthof{\bfseries Conceptual Model}]
	\item[Discoverability] Refers to determining possible actions and the visualization's current state.
	\item[Feedback] Refers to information provided to understand the visualization's state.
	\item[Conceptual Model] Refers to creating a conceptual model of the process illustrated. 
	\item[Affordances] Refers to the relationship between an object and the user's capabilities. 
    \item[Signifiers] Refers to any indicator that communicates appropriate behavior to a user. 
	\item[Mappings] Refers to the relationship between controls and their actions.
	\item[Constraints] Refers to providing constraints that guide actions and ease the interpretation.

\end{description}

This article is organized as follows. \Cref{fsm-description} presents a brief introduction to \fsm{} including its former visualization tool. \Cref{sm-viz} presents the new visualization tools for \ndfa{}s and \pda{}s, and discusses design choices to address the Norman principles. \Cref{viz-debug} presents how visual debugging and validation is performed. \Cref{rw} compares and contrasts with related work. Finally, \Cref{concl} presents concluding remarks and directions for future work.

\section{Brief Introduction to \fsm{}}
\label{fsm-description}

The machine constructors relevant to the work presented in this article are:\\
\begin{tabular}{llrllllllll}
  & \\
  & & \texttt{\textcolor{pakistangreen}{make-ndfa}}:  & K & \sig{} &        & s & F & \delt{}  & \arrow{} & \ndfa{} \\
  & & \texttt{\textcolor{pakistangreen}{make-ndpda}}: & K & \sig{} & \gam{} & s & F & \delt{}  & \arrow{} & \pda{}   \\
  & \\
\end{tabular}\\
\texttt{K} denotes the states, \sig{} denotes the input alphabet, \(\Gamma\) denotes the stack alphabet, \texttt{s}$\in$\texttt{K} denotes the starting state, \texttt{F}$\subseteq$\texttt{K} denotes the final states, and \texttt{$\delta$} denotes the transition relation. The transition relation is represented as a collection of transition rules. For \ndfa{}s, a transition rule is a triple, \texttt{(K $\{\Sigma \cup \{\ep{}\}\}$ K)}, containing a source state, the element to read (possibly none), and a destination state. For \pda{}s, a transition rule is a pair, \texttt{((K $\{\Sigma \cup \{\ep{}\}\}$ $\{\Gamma \cup \{\ep{}\}\}$) (K $\{\Gamma \cup \{\ep{}\}\}$))}, containing a triple and a pair. The triple contains a source state, the element read (possibly none), and the elements to pop off the stack (possibly none). The pair contains a destination state and the elements to push onto the stack (possibly none). Operationally, \pda{}s first pop and then push. Finally, nondeterministic machines accept a word if any possible computation reaches an accepting configuration and reject if all possible computations halt without reaching an accepting configuration. Given that repeated configurations are not explored, \ndfa{}s always decide a language in \fsm{}. As is well-known, on the other hand, a \pda{} may only semidecide a language and, thus, run forever when given a word not in its language.

There are observers to extract machine components: \textcolor{pakistangreen}{\texttt{sm-states}}, \textcolor{pakistangreen}{\texttt{sm-sigma}}, \textcolor{pakistangreen}{\texttt{sm-start}}, \textcolor{pakistangreen}{\texttt{sm-finals}}, \textcolor{pakistangreen}{\texttt{sm-rules}}, and \textcolor{pakistangreen}{\texttt{sm-gamma}}. Additionally, the observers \textcolor{pakistangreen}{\texttt{sm-apply}} and \textcolor{pakistangreen}{\texttt{sm-showtransitions}}, given a machine and a word that may or may not be in \texttt{L(M)}, are used to apply the given machine to the given word. The first returns \accept{} if there is a computation that accepts. Otherwise, it returns \reject{} or, in the case of \pda{}s, may run forever. 
The second returns an ordered set of configurations reached by an accepting computation. For \ndfa{}s, each configuration contains the machine's state and the unconsumed input. For \pda{}s, each configuration contains the machine's state, the unconsumed input, and the stack. For instance, the following is a trace produced for an \ndfa{}:
\begin{alltt}
     (((a b a) S)  ((b a) A)  ((a) C)  (() E)  (() S)  accept)
\end{alltt}
The machine starts in \texttt{S} with \texttt{(a b a)} to be read. At each step, it reads the first (from the left) unread element and changes state. After 3 steps, the machine is in \texttt{E} with no unread input. It then nondeterministically, consuming no input, transitions to \texttt{S}. Given that \texttt{S} is a final state and the input word is entirely consumed, the machine accepts.

A programmer may generate an image for a given machine's transition diagram using \textcolor{pakistangreen}{\texttt{sm-graph}}. In the returned image, a node represents a state and an edge represents a transition rule. The starting state is denoted by a green circle. A final state is denoted by a double black circle. If the starting state is also a final state then it is denoted using a double green circle. All other states are denoted using a single black circle. The label on an edge denotes the actions taken: the element that is read (possibly none) for an \ndfa{} and the element that is read (possibly none), the elements that are pushed, and the elements that are popped for a \pda{}. Such a graphic, illustrated in the next subsection, is generated using \gviz{} \cite{gviz2,gviz1} and its interface with \fsm{} is completely hidden from the programmer. \gviz{} is used because, in general, it produces graph images that are easy to read. The Sugiyama-style graph drawings \cite{Sugiyama} are rendered through a four-pass algorithm that ranks nodes, orders nodes within ranks, finds optimal node-placement coordinates, and makes use of splines \cite{Schumaker1,Schumaker2} for edge drawing \cite{gviz2}.


\subsection{Illustrative Examples}

\begin{figure}
	\begin{enumerate}
		\item Name the machine and specify the alphabet(s)
		\item Write unit tests
		\item Identify conditions that must be tracked as the input is consumed, associate a state with each condition, and determine the start and final states.
		\item Formulate the transition relation
		\item Implement the machine
		\item Run the tests
		\item Design, implement, and test an invariant predicate for each state
		\item Prove L = L(M)
	\end{enumerate}
	\caption{Design recipe for state machines}
	\label{dr}
\end{figure}

To implement machines students follow the steps of the design recipe displayed in \Cref{dr} \cite{PBFLAT}. This section presents an illustrative example for an \ndfa{} and for a \pda{}.

\begin{figure}[t!]
	\captionsetup[subfigure]{justification=centering}
	\begin{subfigure}[b]{\textwidth}
\begin{lstlisting}[language=racket,escapechar=\%]
     #lang fsm
     ;; State Documentation (ci = consumed input)
     ;;   S: ci is empty, starting state
     ;;   A: ci%\textcolor{darkorange}{$\in$}%(ab)%\textcolor{darkorange}{$^*$}%
     ;;   B: the ci%\textcolor{darkorange}{$\in$}%(ab)%\textcolor{darkorange}{$^*$}%a 
     ;;   C: ci%\textcolor{darkorange}{$\in$}%(ab)%\textcolor{darkorange}{$^*$}%b%\textcolor{darkorange}{$^*$}%, final state
     ;;   D: the ci is empty
     ;;   E: ci%\textcolor{darkorange}{$\in$}%ab%\textcolor{darkorange}{$^{\texttt{*}}$}%, final state
     (define ab*-U-ab*b*-ndfa 
       (make-ndfa '(S A B C D E)
                  '(a b)
                  'S
                  '(C E)
                  '((S EMP A) (S EMP D) (A a B) (A EMP C)
                   	(B b A) (C b C) (D a E) (E b E))))
     (check-equal? (sm-apply ab*-U-ab*b*-ndfa '(b a b a a))    'reject)
     (check-equal? (sm-apply ab*-U-ab*b*-ndfa '(a a a))        'reject)
     (check-equal? (sm-apply ab*-U-ab*b*-ndfa '())             'accept)
     (check-equal? (sm-apply ab*-U-ab*b*-ndfa '(a b b b))      'accept)
     (check-equal? (sm-apply ab*-U-ab*b*-ndfa '(a b a b b b))  'accept)
\end{lstlisting}
\caption{The \ndfa{} definition and unit tests in \fsm{}.}  \label{ndfa-impl}
\end{subfigure}
\vfill
\captionsetup[subfigure]{justification=centering}
\begin{subfigure}{\textwidth}
\centering
\includegraphics[scale=.2]{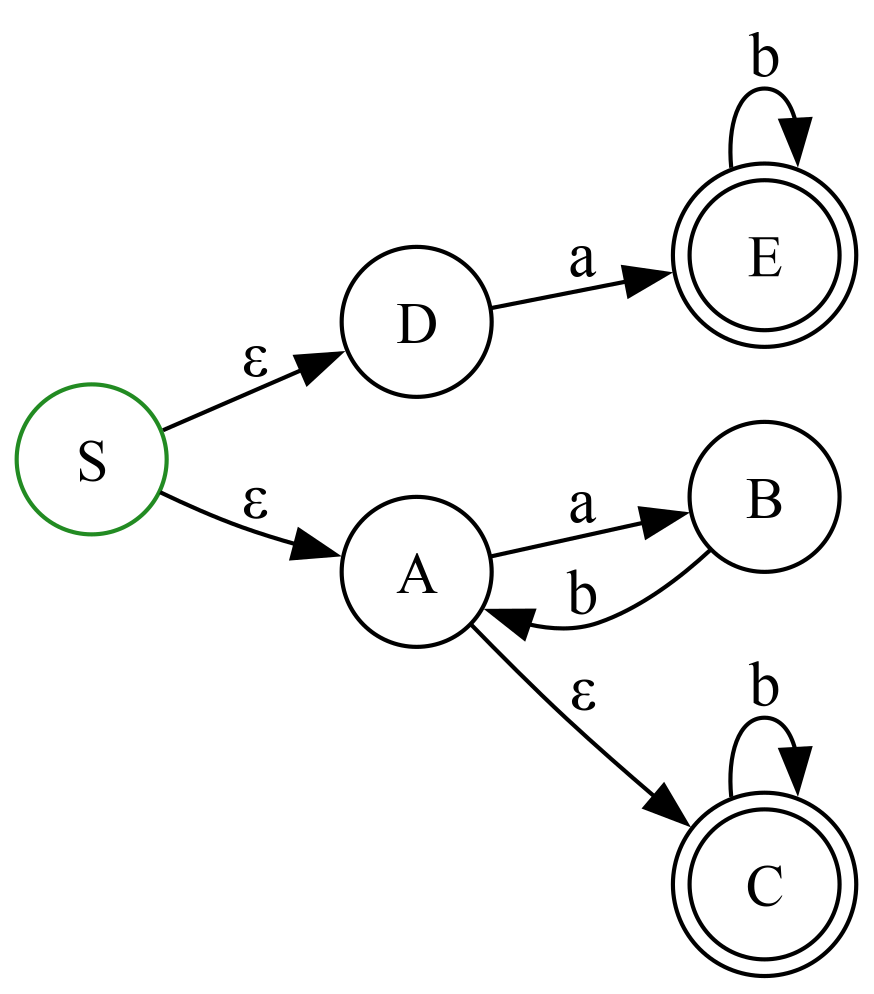}
\caption{The \ndfa{}'s transition diagram.} \label{samplendfa}
\end{subfigure}
\caption{A \ndfa{} for \texttt{L = ab$^{\texttt{*}}$ $\cup$ (ab)$^{\texttt{*}}$b$^{\texttt{*}}$}.}
\end{figure}

\subsubsection{Developing \ndfa{}s}

Consider designing and implementing an \ndfa{} for the language \texttt{ab$^{\texttt{*}}$$\cup$(ab)$^{\texttt{*}}$b$^{\texttt{*}}$}. The \fsm{} implementation is displayed in \Cref{ndfa-impl}. The result for Step 1 of the design recipe is displayed on lines 9 and 11, respectively, the name of the machine and the input alphabet. The unit tests to satisfy Step 2 are displayed on lines 16--20. For Step 3, the state documentation is displayed on lines 2--8. The transition relation developed to satisfy Step 4 is displayed on lines 14--15. Each transition rule is designed assuming that the design of the source state holds and that the element read guarantees that the design of the destination state holds. Step 5 of the design recipe is satisfied by providing \texttt{make-ndfa} the inputs developed in the previous steps. Step 6 is satisfied by running the tests and, if bugs are manifested or tests fail, redesigning.

For Step 7, students develop predicates to validate the role of each state. These predicates require the consumed input and test some property of this given subword. In the interest of brevity, we only show the invariant predicate for \texttt{S}:
\begin{lstlisting}[language=racket,escapechar=\%,numbers=none]
     ;; word -> Boolean    Purpose: To determine if the consumed input is empty
     (define (S-INV ci) (empty? ci))
     
     (check-equal? (S-INV '()) #t)
     (check-equal? (S-INV '(a b a)) #f)
\end{lstlisting}
The predicate clearly communicates that the machine should only be in state \texttt{S} when the consumed input is empty as stated in line 3 in \Cref{ndfa-impl}.

Finally, for Step 8, students must prove that \texttt{L=L(M)}. This is done in two substeps. In the first, students prove by induction that state invariants hold on accepting computations. In the second, they prove the desired equivalence based on state invariants holding on accepting computations. In the interest of brevity, we do not present the complete proofs. For the base case of the induction, the students prove that the invariant for the starting state holds. For \texttt{ab*-U-ab*b*-ndfa}, we note that \texttt{S-INV} holds because the consumed input is empty. For the inductive step, students establish, for each transition rule, the invariant for the destination state holds assuming the invariant for the source state holds. For instance the transition from \texttt{S} to \texttt{A} on empty is verified as follows:
\begin{alltt}
  \underline{(S \ep{} A)}: By the inductive hypothesis, S-INV holds. This means that the 
  consumed input is empty. Using this transition rule does not change the 
  consumed input. Therefore, the consumed input is in (ab)\(\sp{\texttt{*}}\). Thus, A-INV holds.
\end{alltt}
The same is done for each transition that may lead to an accepting configuration. 

To establish the that \texttt{L=L(M)}, students need to prove two lemmas\footnote{By contraposition it is only necessary to prove one lemma \cite{Hurley}. In our experience, however, student maturity with proofs is lacking and, therefore, explicitly proving both lemmas better communicates the equivalence of the languages.}:
\begin{alltt}
     w\(\in\)L \(\Leftrightarrow\) w\(\in\)L(M)     w\(\notin\)L \(\Leftrightarrow\) w\(\notin\)L(M)
\end{alltt}
We illustrate the proof of the first lemma as follows:
\begin{alltt}
 (\(\Rightarrow\)) Assume w\(\in\)L
      w\(\in\)L means that w\(\in\)(ab)\(\sp{*}\)b\(\sp{*}\) or w\(\in\)ab\(\sp{*}\). Given that invariants always hold,
      the machine halts in either C or E after consuming all the input. 
      Thus, w\(\in\)L(M).
 (\(\Leftarrow\)) Assume w\(\in\)L(M)
      This means the machine halts in states C or E after consuming w. 
      Given that invariants always hold, we have that w\(\in\)(ab)\(\sp{*}\)b\(\sp{*}\) or w\(\in\)ab\(\sp{*}\). 
      Thus, w\(\in\)L.
\end{alltt}
The proof for the second lemma is written in a similar fashion.

The transition diagram for \texttt{ab*-U-ab*b*-ndfa} is displayed in \Cref{samplendfa}. Observe that the machine must make a nondeterministic choice in \texttt{S}, the starting state, to move to \texttt{A} or \texttt{D}. This is a precise example of where some students get confused. They naturally ask \emph{How does the machine make this choice?} In part, the new dynamic visualization tool has been developed to provide an understandable answer to students.

\begin{figure}[t!]
\captionsetup[subfigure]{justification=centering}
\begin{subfigure}{\textwidth}
\begin{lstlisting}[language=racket,escapechar=\%]
#lang fsm
;; State Documentation
;;   S: (append ci stack) has equal number of a's and b's
(define P (make-ndpda '(S)
                      '(a b)
                      '(a b)
                      'S
                      '(S)
                      '(((S a EMP)   (S (b))) ((S a (a)) (S EMP))   ;; pop matching letter
                        ((S b (b)) (S EMP))   ((S b EMP) (S (a))))));; push opposite letter
(check-equal? (sm-apply P '())            'accept)
(check-equal? (sm-apply P '(a b b))       'reject)
(check-equal? (sm-apply P '(a))           'reject)
(check-equal? (sm-apply P '(b a b))       'reject)
(check-equal? (sm-apply P '(b a a b))     'accept)
(check-equal? (sm-apply P '(a b b a a b)) 'accept)

(define (S-INV ci stack)
  (let* [(ci+stack (append ci stack))
         (as (filter (lambda (s) (eq? s 'a)) ci+stack))
         (bs (filter (lambda (s) (eq? s 'b)) ci+stack))]
    (= (length as) (length bs))))
    
(check-equal? (S-INV '(b a) '(b b b)) #f)
(check-equal? (S-INV '(a) '(a)) #f)
(check-equal? (S-INV '() '()) #t)
(check-equal? (S-INV '(a a b) '(b)) #t)
\end{lstlisting}
\caption{The \pda{} implementation.} \label{pda-impl}
\end{subfigure}
	\vfill
	\captionsetup[subfigure]{justification=centering}
	\begin{subfigure}{\textwidth}
       \centering
		\includegraphics[scale=.2]{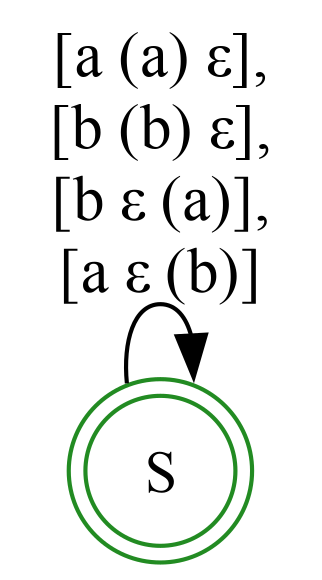}
		\caption{The \pda{}'s transition diagram.}
		
		\label{samplepda}
	\end{subfigure}
	\caption{A \pda{} for \texttt{L = \{w|w has an equal number of as and bs\}}.}
\end{figure}

\subsubsection{Developing \pda{}s}

Consider implementing a \pda{} for the language of all words that have an equal number of \texttt{a}s and \texttt{b}s. The machine developed using the design recipe from \Cref{dr} is displayed in \Cref{pda-impl}. The design idea is for the machine to nondeterministically decide the action taken when a word element is read: push the opposite letter onto the stack or pop the same letter from the stack. The machine's name, input alphabet, and stack alphabet to satisfy Step 1 are displayed on lines 4, 5, and 6. The unit tests developed to satisfy Step 2 are displayed on lines 11--16. The state documentation developed for Step 3 is displayed on lines 2--3. A single state is needed that represents that the consumed input and the stack combined have an equal number of \texttt{a}s and \texttt{b}s. Step 4 is satisfied by the transition relation displayed on lines 9--10. As with \ndfa{}s, each transition is developed assuming that the role of the source state and the actions taken by the transition guarantee that the role of the destination state is satisfied. The implementation using the \pda{} constructor satisfies Step 5 and Step 6 is satisfied by running the program and discovering no errors and no failed tests. The transition diagram for the constructed \pda{} is displayed in \Cref{samplepda}.

To satisfy Step 7, a single invariant predicate is needed. Such a predicate requires the consumed input and the stack value. The implementation of this predicate for state \texttt{S} is displayed in lines 18--22 in \Cref{pda-impl}. Finally, to satisfy Step 8, students prove that state invariant predicates hold for computations that lead to accept and establish the same two lemmas as for \ndfa{}s to conclude that the machine's language is correct. In the interest of brevity, we omit illustrating these proofs.

\begin{figure}[t!]
	\centering
	\begin{subfigure}{.8\textwidth}
		\captionsetup{justification=centering}
		\includegraphics[width=\textwidth]{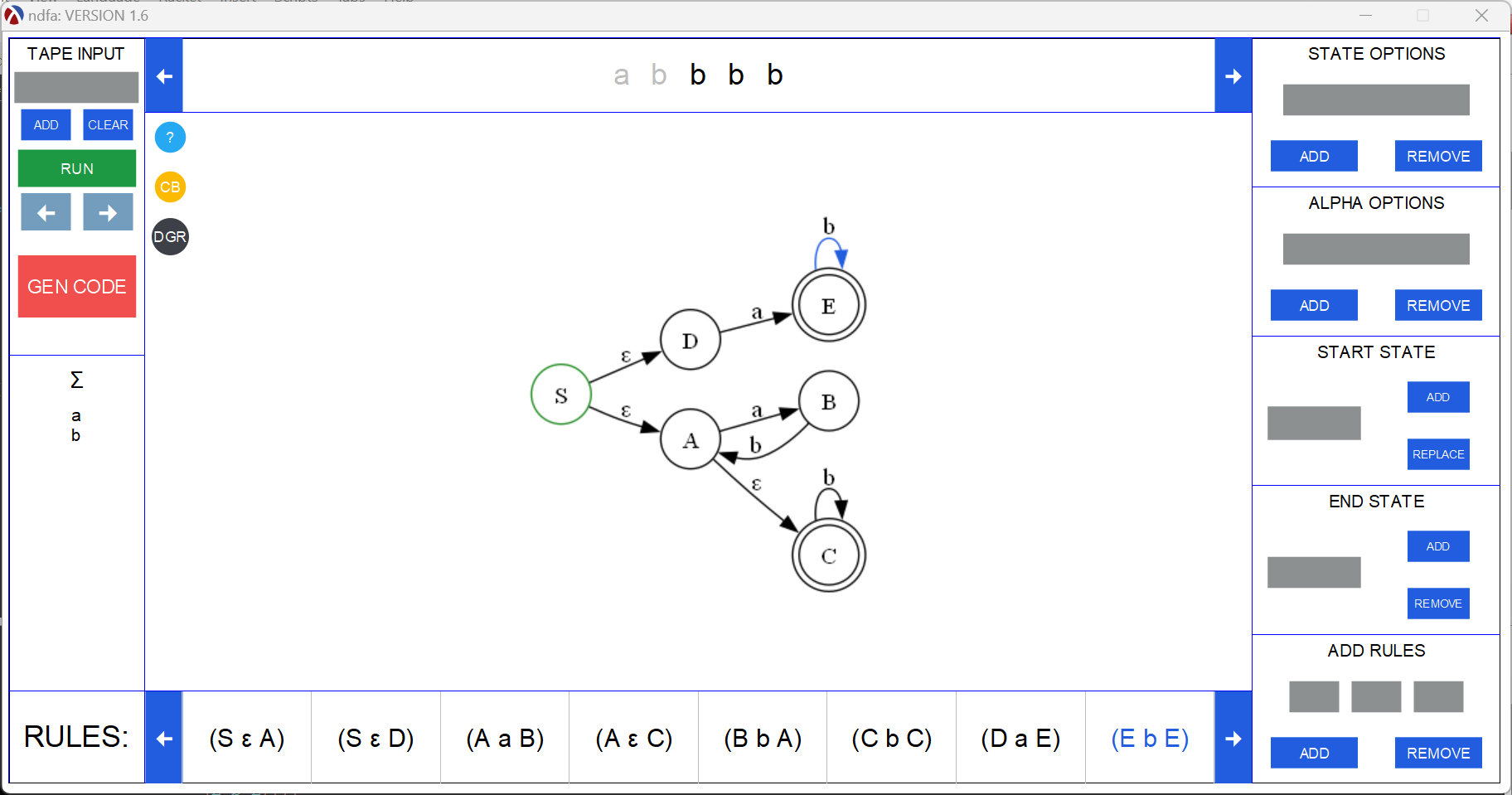}
		\caption{\texttt{sm-visualize} using the \ndfa{} designed in \Cref{ndfa-impl}}
		\label{sm-visualize-ndfa}
	\end{subfigure}
	\vfill
	\begin{subfigure}{.8\textwidth}
		\captionsetup{justification=centering}
		\includegraphics[width=\textwidth]{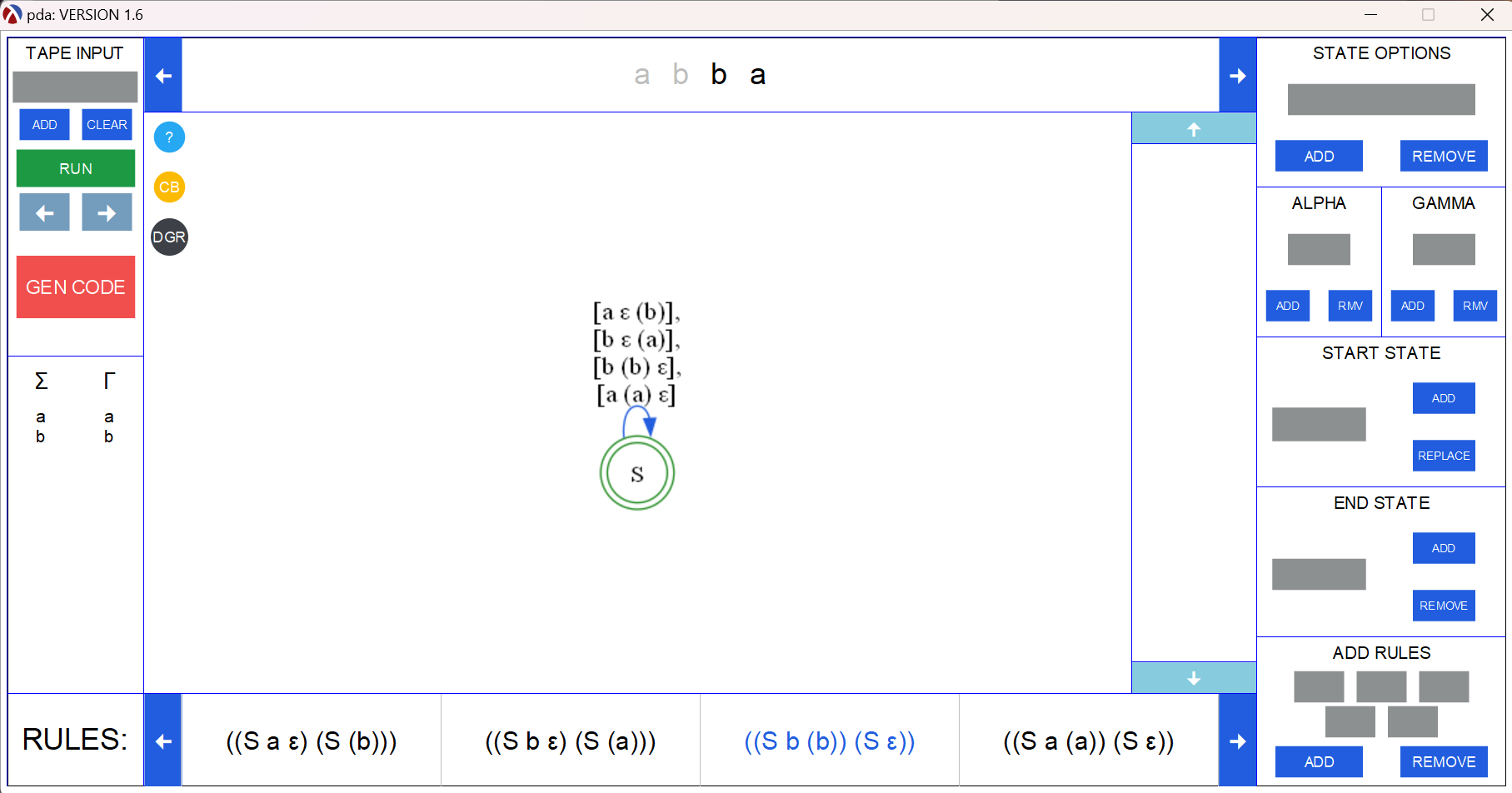}
		\caption{\texttt{sm-visualize} using the \pda{} designed in \Cref{pda-impl}}
		\label{sm-visualize-pda}
	\end{subfigure}
    \caption{The former \fsm{} visualization tool.} \label{sm-visualize-tool}
\end{figure}

\subsection{\fsm{}'s Original Visualization Tool}

\fsm{}'s first visualization tool allows users to visualize the application of \ndfa{}s and \pda{}s \cite{fsm-viz}. \Cref{sm-visualize-tool} displays snapshots of the dynamic visualization for the \ndfa{} displayed in \Cref{ndfa-impl} and for the \pda{} displayed in \Cref{pda-impl}. This tool allows the user to provide an input word and navigate the visualization forward and backwards (left column), to edit the machine (right column), and displays the machine's configuration and highlights the last rule used (center column). It allows the user to provide state invariant predicates. When an invariant predicate, \texttt{Q-INV}, is provided for a state, \texttt{Q}, its node is highlighted in green when the machine transitions into it and \texttt{Q-INV} holds. If the \texttt{Q-INV} does not hold then \texttt{Q}'s node is highlighted in red.

Although useful, this tool has proven lacking for several reasons. Students rarely, if ever, edited machines in the visualization tool. Instead, they edited their program and ran the tool again. Invariably, students expressed that they would prefer a larger image for the transition diagram displayed. This led to our realization that zooming facilities would be more useful than machine editing facilities. Students also regularly complained that finding configurations where a state invariant failed required ``endless'' clicking forward or backwards. This lead to our realization that a dynamic visualization tool needed a feature to jump in one step to the next or the previous configuration where an invariant fails. Finally, the tool offered no assistance in understanding why a word is rejected, because it only informs the user that the word is not in the machine's language. This made us aware that a dynamic counterpart for \fsm{} computation graphs \cite{fsm-compgraphs}--a static visualization tool that helps explains why words are rejected--is needed. .

\section{The New Dynamic Visualization Tool}
\label{sm-viz}

\subsection{Design Idea}

To address the shortcomings of \fsm{}'s first dynamic visualization tool for state machines, we have developed a new tool for the stepwise execution of \ndfa{}s and \pda{}s. This tool has the following characteristics:
\begin{itemize}
  \item[$\circ$] Larger space for the main visualization graphic (i.e., the transition diagram)
  \item[$\circ$] One step to move the visualization forwards, backwards, to the beginning, and to the end
  \item[$\circ$] Zooming capabilities
  \item[$\circ$] Scrolling for large input words
  \item[$\circ$] Trace all computations without repeated configurations
  \item[$\circ$] When a \pda{} accepts, trace the stack for a single accepting computation
  \item[$\circ$] Highlighting of accepting computation transitions in green
  \item[$\circ$] Highlighting of rejecting computation transitions in violet
  \item[$\circ$] Optionally, a dead/trap state may be added to guarantee that all computations consume the entire input word (as done by \dfa{}s)
  \item[$\circ$] A user-adjustable cut off threshold for the number of steps taken by a \pda{} computation
  \item[$\circ$] Rendering states where \pda{} computations are cut off in gold
  \item[$\circ$] Optionally, visualize the value of state invariant predicates by coloring states
      \begin{itemize}
        \item[$\star$] Green for holding
        \item[$\star$] Red for not holding
        \item[$\star$] Display invariant values only for accepting computations
        \item[$\star$] Bicolor state rendering when for at least two different accepting computations reaching the same state the invariant holds for one computation and fails for the other
      \end{itemize}
  \item[$\circ$] One step to move the visualization to the next or previous configuration, if any, in which a state invariant predicate fails
\end{itemize}
One of the primary goals is to offer a tool that presents a low extraneous cognitive load to students. That is, the tool ought to be intuitive and not shift focus from designing/debugging/validating \ndfa{}s and \pda{}s to learning how to use the tool. The first four bullets address common requests made by students. Real estate inside the visualization's frame is made available by eliminating the unused facilities to edit machines within the visualization tool and by changing the interface based on mouse clicking to an interface that allows keystrokes and mouse clicking.

The fifth through eighth bullets aim to help students understand how and why nondeterminism leads to multiple computations on the same input. In addition, it aims to help students understand why a word is accepted/rejected. The use of color highlights that there can be multiple accepting and multiple rejecting computations. Finally, inspired by the work done with computation graphs \cite{fsm-compgraphs}, computations that reach a configuration that has already been explored are no longer explored. Such an exploration would only repeat the same computations performed when the configuration is visited the first time.

The ninth bullet aims to alleviate the anxiety some students feel when a computation does not consume the entire input. This anxiety, as best as we can tell, stems from the study of \dfa{}s before studying nondeterministic machines. As is well-known, a \dfa{} must consume all its input before halting. Despite being told the contrary, some students seem to cling to the belief that a computation must consume all its input. To combat this belief, the new dynamic visualization tools allow for an optional dead state to be added to \ndfa{}s and \pda{}s. Instead of halting before entirely consuming the input word and/or emptying the stack, the machine transitions to the added dead/trap state where the remaining part of the input is consumed and/or the stack is emptied before halting and rejecting.

The tenth and eleventh bullets aim to help students understand that a \pda{} may perform infinite computations. They can increase the cutoff threshold to determine if allowing for more steps leads all computations to terminate. 

Finally, the last two bullets aim to assist students in the verification process. Students validate their design by observing if state invariant predicates hold before attempting a proof. The user may provide an optional list of pairs containing a state and its invariant predicate. As the visualization progresses, transitioned-into states in accepting computations for which the invariant holds are rendered green, transitioned-into states in accepting computations for which the invariant fails are rendered in red, and, otherwise, states are rendered without the use of color. We note that for \pda{}s it is possible for a state invariant predicate to hold for one accepting computation and not hold for a different accepting computation. In such a case, the state's rendering is bicolor. Finally, the goal behind the last bullet is to reduce the amount of time needed to find configurations in which a state invariant predicate fails.

\subsubsection{Graphical Interface}

The visualization frame is divided into three vertically-aligned parts: the main visualization graphic, informative messages, and instructions. The first two are dynamic and change as the visualization advances or steps back. The third is static and remains constant through all visualization steps.

The main visualization graphic always displays the machine's transition diagram color-coded to easily identify the possible computations (without shared configurations). At each step, the edge denoting the last transition used in every computation that has not ended is highlighted in color: green for accepting computations, if any, and violet for all other computations. Two tones of green are used for accepting computations. The darker tone indicates the accepting computation that is tracked exemplifying why the word is accepted. This is especially important for visualizing \pda{}s: the informative messages display the stack value of the tracked accepting computation. After the last element is consumed by an accepting computation, the visualization displays an informative message stating that there is an accepting computation. When all computations halt in a non-accepting configuration, an informative message is displayed stating that the given word is rejected. It is, of course, possible for a \pda{} to only semidecide a language and run forever when a word is not in the machine's language. To prevent infinite executions, as mentioned above, a user-adjustable cutoff threshold for the maximum number of steps allowed is available. Computations are discontinued when this threshold is reached and the state in which the threshold is reached in rendered in gold. In this manner, users are made aware that the visualized computation is incomplete and that there exists computations that were not allowed to terminate. 

The middle part of the visualization consists of informative messages, which provide context to understand the state of the main visualization graphic after the last step taken. The informative messages always align the input word and the consumed input. The consumed input is faded out and the unconsumed input is rendered in black in the message displaying the input word. The consumed input message is always rendered in black and always corresponds to the faded portion in the input word. When visualizing \pda{}s, the stack is only displayed for, if any, one accepting computation that is being traced (highlighted using the darker tone of green in the main visualization graphic)\footnote{The decision to not display the stack for all possible computations is based on the cluttering effect multiple stacks can have on a visualization, thus, reducing readability and visual appeal.}. Finally, the informative messages also include the number of possible computations without a shared configuration.

The bottommost component of the visualization presents the instructions to interact with the visualization as key images underlined with a message indicating their functionality. The user may use the corresponding keystrokes or click on the instruction icons to interact with the visualization. 

We now outline how our interface's design addresses the Norman principles of effective design:
\begin{description}[leftmargin=!,labelwidth=\widthof{\bfseries Conceptual Model}]
\item[Discoverability] The use of color in the main visualization graphic and the informative messages help the user understand the current visualization state. The instruction buttons help the user understand how to interact with the visualization tool.
\item[Feedback] Each step is rendered with informative messages to help interpret the main visualization.
\item[Conceptual model] By stepping through all computations (without shared configurations), the user is assisted in building a conceptual model for nondeterministic execution.
\item[Affordances] The description of the keyboard buttons provide the user with a clear understanding of the expected result upon using them.
\item[Signifiers] Arrow color highlighting is used to signal the current state of each accepting and rejecting computation. Node coloring is used to signal where computations are cutoff and whether invariant properties hold for accepting computations. Finally, the informative messages inform the user how much of the input word has been read.
\item[Mappings] The action associated with a key is unchanged during a visualization session.
\item[Constraints] Users may only use the small number of keys displayed in the instructions. Thus, restricting how to interact with the visualization. Furthermore, limits on zooming and dragging guarantee that a part of the main visualization graphic is always visible.
\end{description}

\subsection{Illustrating the \ndfa{} Visualization Tool}

\begin{figure}[t!]
	\centering
	\begin{subfigure}{\textwidth}
		\centering
		\captionsetup{justification=centering}
		\includegraphics[scale=0.9]{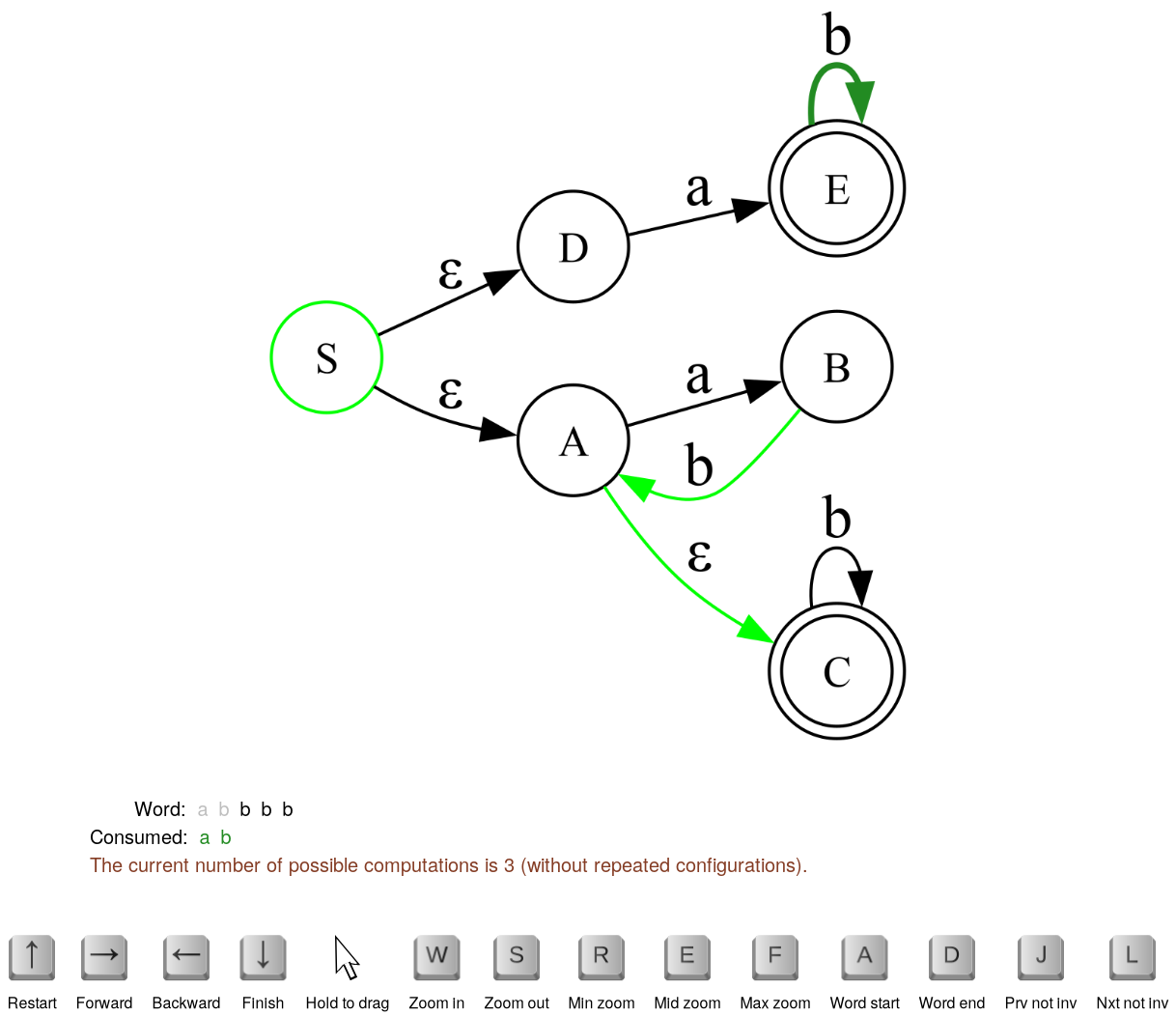}
		\caption{All computations after 2 steps.}
		\label{samplendfaviz}
	\end{subfigure}
	\begin{subfigure}{\textwidth}
        \centering
		\captionsetup{justification=centering}
		\includegraphics[scale=0.9]{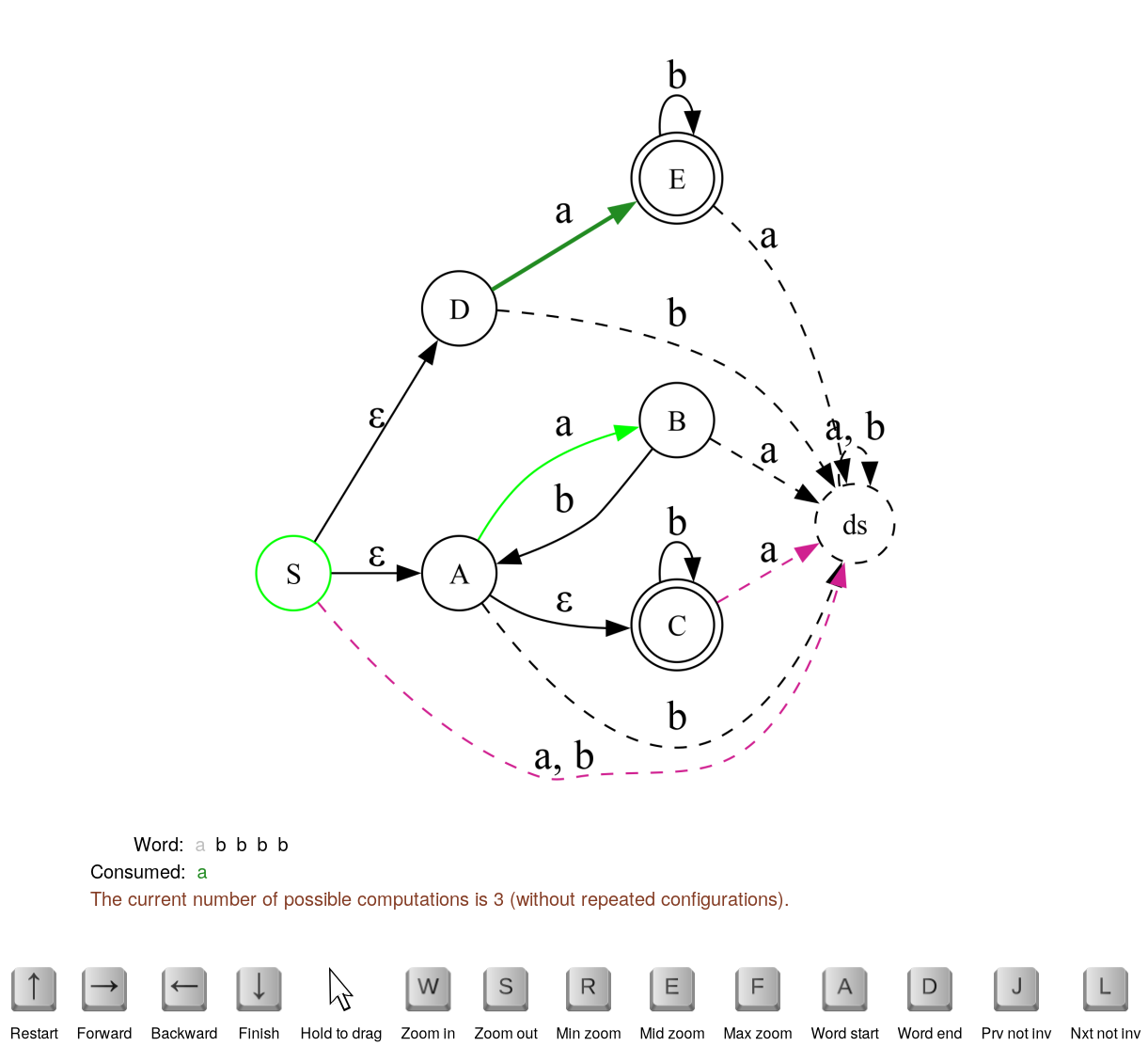}
		\caption{All computations after 1 step with an added dead state.}
		\label{add-deadndfa}
	\end{subfigure}
	\caption{Different views using the new \ndfa{} visualization tool.}
	\label{altvizviews}
\end{figure}

Consider applying \texttt{ab*-U-ab*b*-ndfa} from \Cref{ndfa-impl} to \texttt{(a b b b b)}. Upon launching the tool, the transition diagram is displayed with the two \ep{}-transitions from the starting state, \texttt{S}, highlighted in green indicating that, so far, there are two accepting computations. After consuming \texttt{(a b)}, there are three different computations: one is in \texttt{A}, one is in \texttt{C}, and one is in \texttt{E}. The visualization's state is displayed in \Cref{samplendfaviz}. Observe that the edges denoting the transitions \texttt{(B b A)}, \texttt{(A \ep{} C)}, and \texttt{(E b E)} are highlighted in green. This informs the user that they are part of different accepting computations. In addition, \texttt{(E b E)} is highlighted in a darker tone of green to inform the user that it is part of the computation that is tracked to exemplify why the word is accepted. The informative messages indicate that \texttt{ab} has been consumed and that there are three possible computations.

By stepping back through the computation and following the green arrows in reverse, a student can derive the three computations:
\begin{alltt}
   (S abbbb) \step{} (D abbbb) \step{} (E bbbb) \step{} (E bbb)

   (S abbbb) \step{} (A abbbb) \step{} (B bbbb) \step{} (A bbb)

   (S abbbb) \step{} (A abbbb) \step{} (B bbbb) \step{} (A bbb) \step{} (C bbb)
\end{alltt}
Based on our experience, this is an effective means to foster understanding of nondeterministic behavior among beginners in \flatt{}.

\Cref{add-deadndfa} displays the same computation after only consuming \texttt{a} and using the addition of, \texttt{ds}, a dead/trap state. Dashed edges into this added state are used to indicate that these represent transitions that are not listed in the machine's transition relation and that are added just to enable all computations to consume the entire input word. The main graphic displays four highlighted edges indicating a step in four different computations: one into \texttt{B}, one into \texttt{E}, and two in \texttt{ds}. Observe that the two computations into \texttt{ds} end in the same machine configuration and, therefore, are now the same computation. This is why the informative messages indicate that there are only 3 computations without a shared configuration. 

\begin{figure}[t!]
	\centering
		\includegraphics[scale=1]{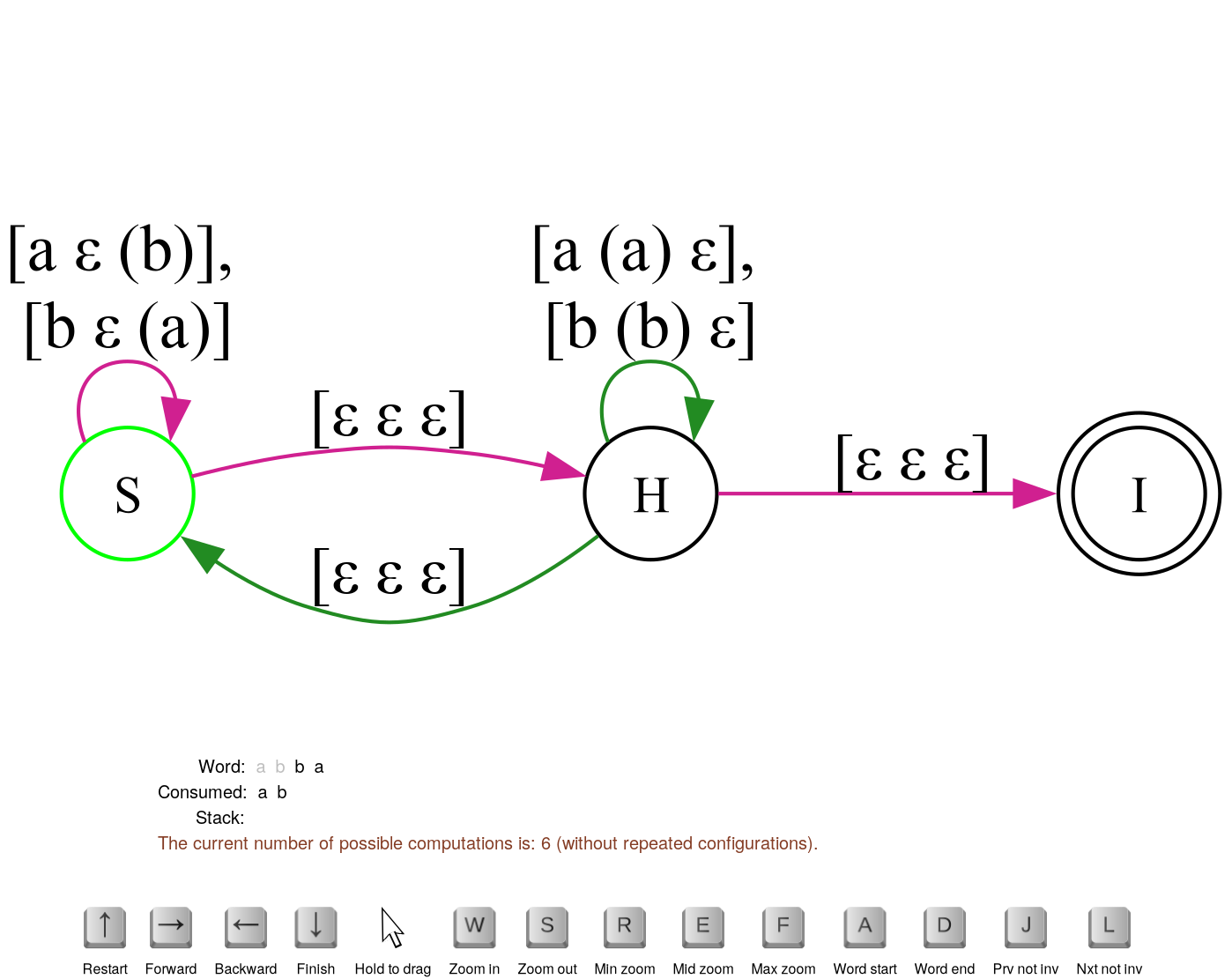}
		\caption{The tool using a \pda{} after 2 steps.}
		\label{samplepdaviz}
\end{figure}

\subsection{Illustrating the \pda{} Visualization Tool}

We consider the application of a student-defined \pda{} for the language containing words with equal number of \texttt{a}s and \texttt{b}s. The state of the visualization after two steps is displayed in \Cref{samplepdaviz}. Note that the starting state, outlined in green, is \texttt{S}. The main visualization graphic immediately communicates that there are at least 5 computations: 2 that accept and 3 that reject. The informative messages, however, indicate that there are 6 computations without shared configurations. The student can trace back to discover the computations. There are 3 computations that consume \texttt{(a b)} by pushing \texttt{b} and then pushing \texttt{a}:
\begin{alltt}
	(S (a b b a) ()) \step{} (S (b b a) (b)) \step{} (S (b a) (a b))
	
	(S (a b b a) ()) \step{} (S (b b a) (b)) \step{} (S (b a) (a b)) \step{} (H (b a) (a b))
	
	(S (a b b a) ()) \step{} (S (b b a) (b)) \step{} (S (b a) (a b)) \step{} (H (b a) (a b))
	                 \step{} (I (b a) (a b))
\end{alltt}
The first remains in \texttt{S}, the second nondeterministically without consuming any input or altering the stack transitions to \texttt{H}, and the third in the same manner transitions from \texttt{H} to \texttt{I}. There are 3 computations that consume \texttt{a} in \texttt{S}, nondeterministically without consuming any input or altering the stack transition to \texttt{H}, and in \texttt{H} consume \texttt{b} and pop \texttt{b} from the stack:
\begin{alltt}
	(S (a b b a) ()) \step{} (S (b b a) (b)) \step{} (H (b b a) (b)) \step{} (H (b a) ())
	
	(S (a b b a) ()) \step{} (S (b b a) (b)) \step{} (H (b b a) (b)) \step{} (H (b a) ())
	                 \step{} (I (b a) ())
	
	(S (a b b a) ()) \step{} (S (b b a) (b)) \step{} (H (b b a) (b)) \step{} (H (b a) ()) 
	                 \step{} (S (b a) ())
\end{alltt}
The first remains in \texttt{H}, the second nondeterministically without consuming any input or altering the stack transitions to \texttt{I}, and the third in the same manner transitions to \texttt{S}. Some of these computations may continue by looping nondeterministically between \texttt{H} and \texttt{S}. We point out to students that these computations generate a configuration that has been explored and, therefore, are no longer explored by the visualization tool.

Observe that there are two edges highlighted with the darker tone of green. This is a feature that indicates that both edges are used by the accepting computation being tracked. That is, the tracked accepting computation reads and pops a \texttt{b} and nondeterministically moves to \texttt{S}. 

\begin{figure}[t!]
\begin{subfigure}{.60\textwidth}
\captionsetup{justification=centering}
\begin{lstlisting}[language=racket,escapechar=\%,numbers=none]
(define numb>numa 
  (make-cfg 
    '(S A)
    '(a b)
    '((S -> b)      (S -> AbA)
      (A -> AaAbA)  (A -> AbAaA)
      (A -> EMP)      (A -> bA))
    'S))

(check-equal? 
  (grammar-derive numb>numa '(a b))
  "(a b) is not in L(G).")
(check-equal? 
  (grammar-derive numb>numa '(a b a))
  "(a b a) is not in L(G).")
(check-equal? 
  (grammar-derive numb>numa '(a a a a a))
  "(a a a a a) is not in L(G).")
(check-equal? 
  (grammar-derive numb>numa '(b b b))
  '(S -> AbA -> bA -> bbA -> bbbA -> bbb))
\end{lstlisting}
\caption{\fsm{} context-free grammar.} \label{cfg}
    \end{subfigure}
	\begin{subfigure}{.39\textwidth}
		\captionsetup{justification=centering}
		\includegraphics[width=\textwidth]{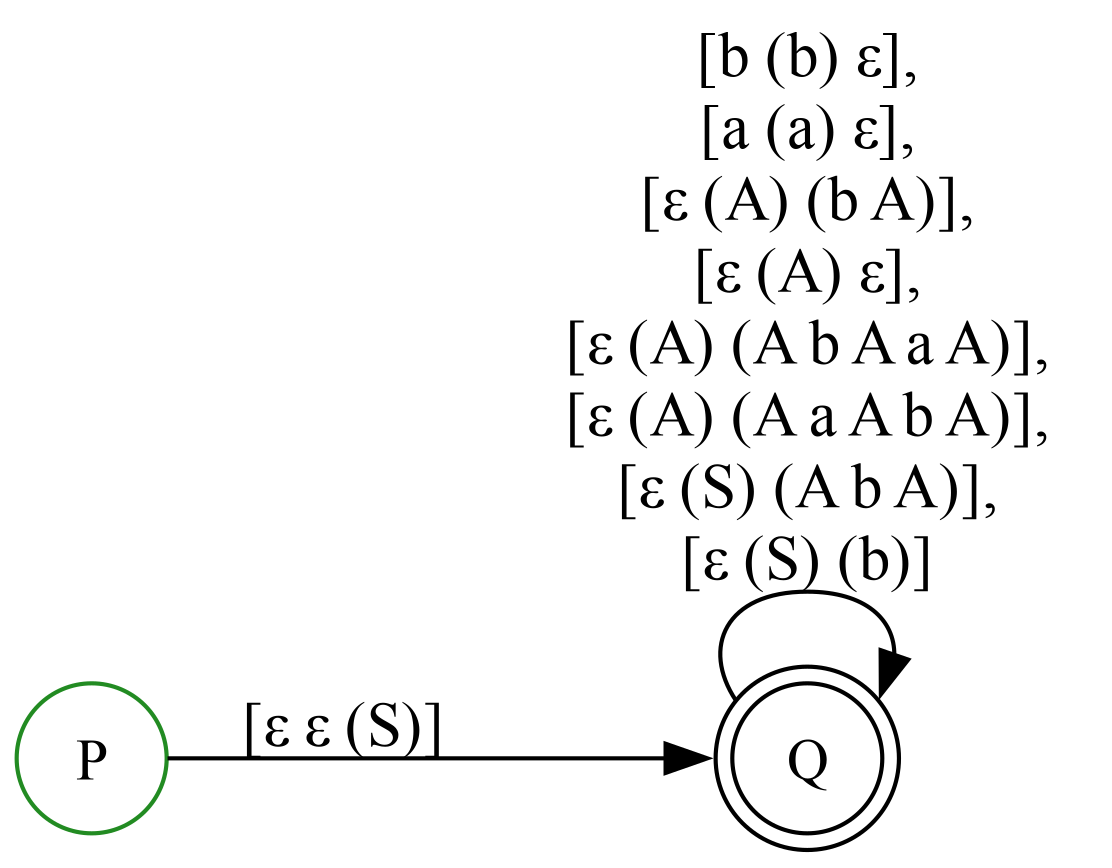}
		\caption{Resulting \pda{} called \texttt{numb>numa-pda}.}
		\label{pda-res}
	\end{subfigure}
	\label{pda-viz}
	\caption{Context-free grammar for L=\{w$|$w$\in$\{a b\}$^*$ $\wedge$  w has more \texttt{b}s than \texttt{a}s\} and resulting \pda{}.}
\end{figure}

\begin{figure}[t!]
\centering
\includegraphics[scale=0.8]{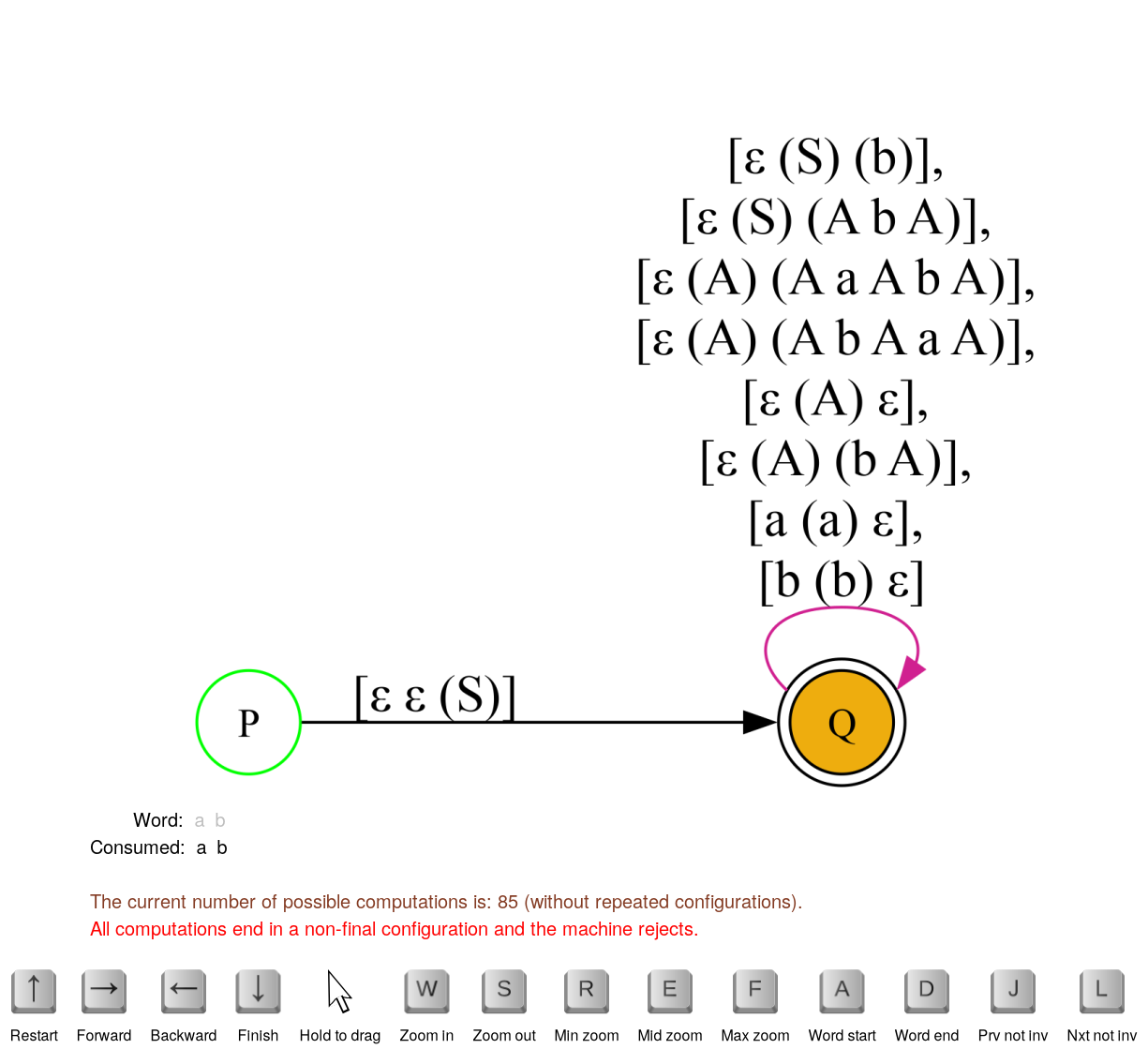}
\caption{A \pda{} visualization snapshot with cut off computations.} \label{cutoff-pda}
\end{figure}

To illustrate how cutoff computations are rendered, consider the result of transforming the context-free grammar displayed in \Cref{cfg} to the \pda{} displayed in \Cref{pda-res}. The resulting \pda{} is the result of applying the transformation algorithm students learn and implement as part of the course \cite{PBFLAT}. It is natural for students to test the resulting \pda{} using the words in the grammar's tests. One such test is:
\begin{lstlisting}[language=racket,escapechar=\%,numbers=none]
     (check-equal? (sm-apply numb>numa-pda '(a b)) 'reject)
\end{lstlisting}
The application of \texttt{numb>numa-pda} results in an infinite computation and, therefore, the test never returns. Invariably, this puzzles students because they have not assimilated that \pda{}s may only semidecide a language. The problem here is that there are computations that use the following rules indefinitely:
\begin{alltt}
          (A -> AaAbA)     (A -> AbAaA)
\end{alltt}
The new \pda{} dynamic visualization tool may be used to help students understand why this occurs. Consider the visualization state displayed in \Cref{cutoff-pda} using a cut off threshold of 10. The informative messages communicate that there are 85 cut off computations. Given that the input word is of length 2, it is unequivocal that nondeterministic transitions are at play. After examining the transitions on \texttt{Q}, where the computations are cut off, it becomes clear that transitions \texttt{((Q \ep{} (A)) (Q (A a A b A)))} and \texttt{((Q \ep{} (A)) (Q (A b A a A)))} may be used an arbitrary number of times without creating a repeated machine configuration. Thus, there are computations that are infinite and that is why the test never completes.

\section{Visual Debugging and Verification}
\label{viz-debug}

The presented visualization tool is also be used to debug the implementation of \ndfa{}s, \pda{}s, and state invariant predicates. In this section, we present a debugging example for each type of machine.

\subsection{Debugging \ndfa{}s}

Consider the following student-developed \ndfa{} for \lang{}:
\begin{center}
\includegraphics[scale=0.2]{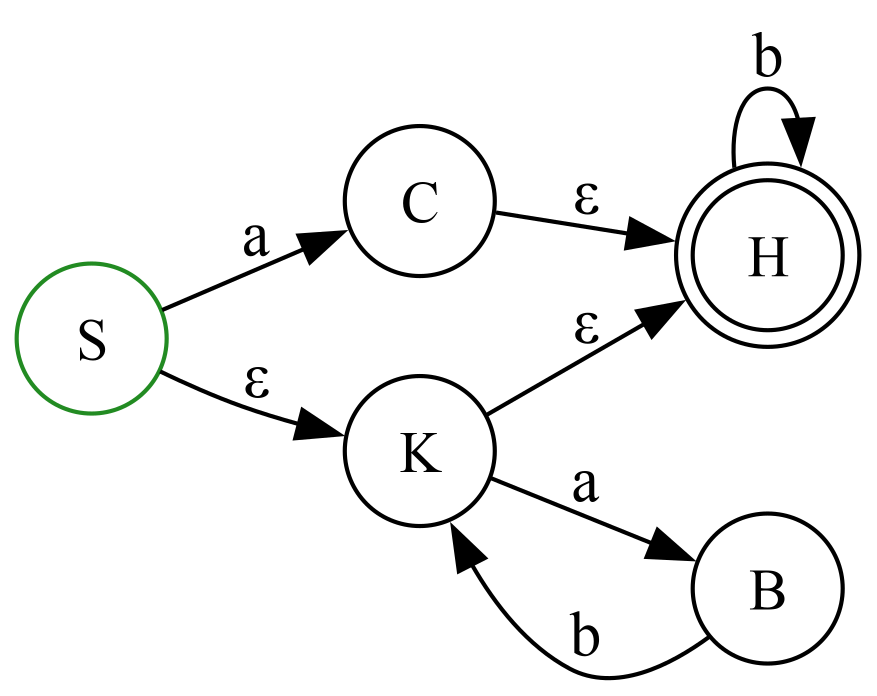}
\end{center}
In addition, consider two of state invariant predicates developed by the student:
\begin{center}
\begin{lstlisting}[language=racket,escapechar=\%,numbers=none]
     ;; word -> Boolean
     ;; Purpose: Determine last letter is not an a
     (define (B-INV a-word)
       (and (not (empty? a-word)) (not (eq? (last a-word) 'a))))

     ;; word -> Boolean
     ;; Purpose: Determine empty or last letter is a or b
     (define (H-INV a-word)
       (or (empty? a-word) (eq? (last a-word) 'a) (eq? (last a-word) 'b)))
\end{lstlisting}
\end{center}
The \flatt{} professor advises the student that:
\begin{itemize}      
  \item \texttt{B-INV} asserts that the consumed input does not end with an \texttt{a}
      
  \item \texttt{H-INV} always holds regardless of what the consumed input is
\end{itemize}

\begin{figure}
	\centering
	\includegraphics[scale=0.9]{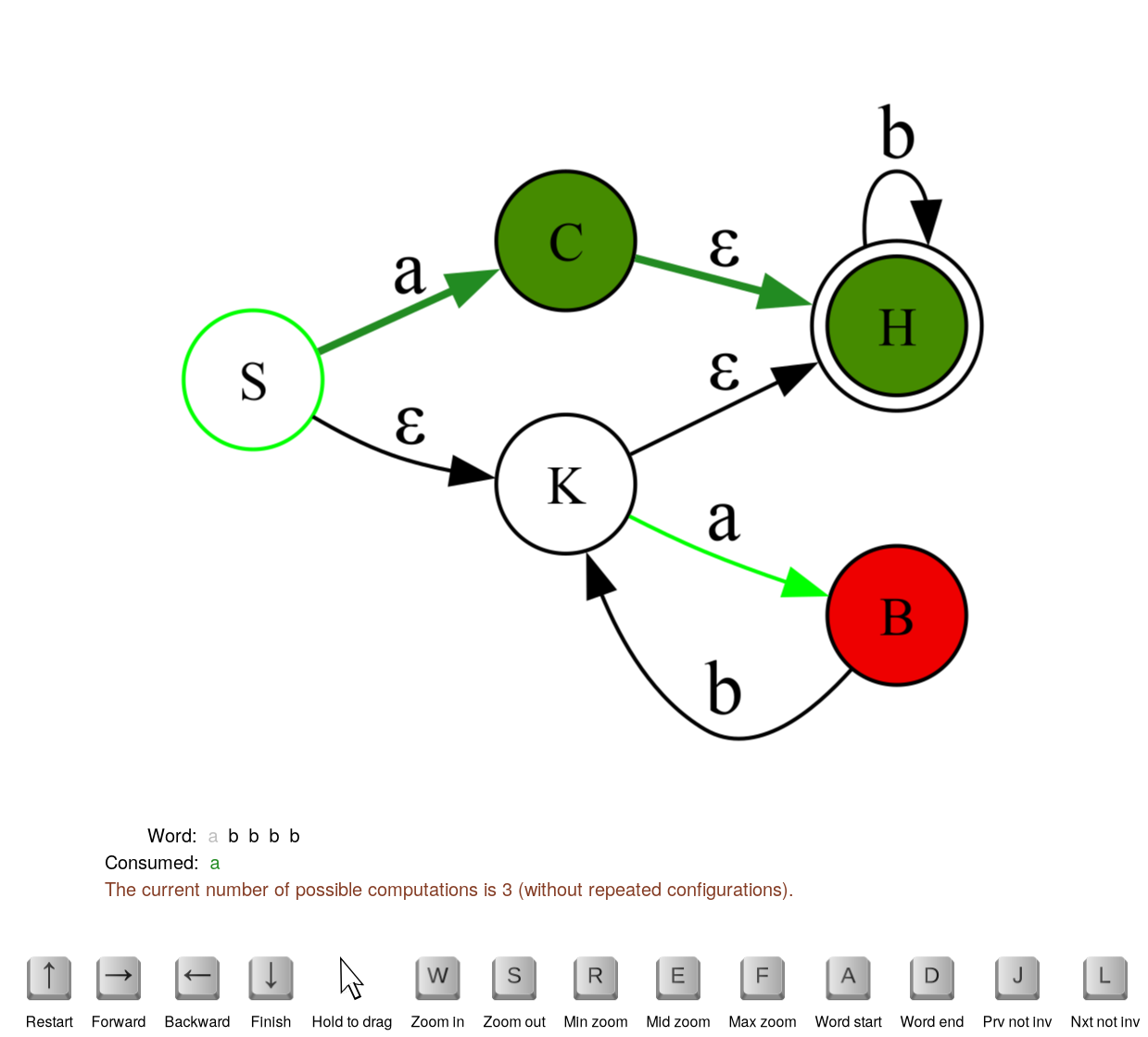}
	\caption{Debugging \ndfa{} state invariant predicates.}
	\label{invs-ndfa}
\end{figure}

The student uses the new dynamic visualization tool to debug their design. \Cref{invs-ndfa} displays a snapshot of the visualization tool given the word \texttt{\quot{}(a b b b b)}. After consuming the first \texttt{a}, the student can observe that \texttt{B-INV} does not hold and revisits to Step 7 of the design recipe. The graphic clearly shows that the last element consumed upon reaching \texttt{B} is an \texttt{a}. Therefore, either the transition into \texttt{B} is buggy or \texttt{B-INV} is buggy. In this case, \texttt{B-INV} must be fixed by removing the \texttt{not} to test if the consumed input ends with \texttt{a}. We note that the graphic has two edges highlighted in the darker tone of green to indicate that the accepting computation tracked uses both transitions.
 
Running the visualization again reveals that \texttt{B-INV} holds. The tool, however, does not detect the bug in \texttt{H-INV}. This drives home the point that running the visualization tool is no substitute for unit tests and proving the correctness of the machine. In this case, upon pressing the student to write unit tests for \texttt{H-INV} the bug is revealed:
\begin{lstlisting}[language=racket,escapechar=\%,numbers=none]
     (check-equal? (H-INV '(a b a)) #t)
\end{lstlisting}
The student observes that \texttt{H-INV} should not hold, because there is no transition into \texttt{H} on an \texttt{a} and proceeds to correct it.

\subsection{Debugging \pda{}s}

Consider the student-developed \pda{} for \texttt{L = \{a$^{\texttt{i}}$b$^{\texttt{j}}$|i\(\leq\)j\(\leq\)2i\}}:
\begin{center}
\includegraphics[scale=0.2]{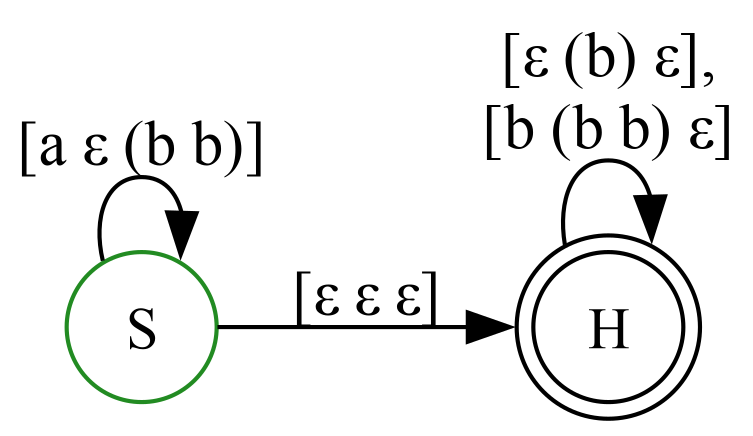}
\end{center}
In addition, the student has developed the following state invariant predicates:
\begin{lstlisting}[language=racket,escapechar=\%,numbers=none,mathescape=true]
;; word (listof %\textcolor{darkgreen}{$\Gamma$}%) %\textcolor{darkgreen}{$\rightarrow$}% Boolean
;; Purpose: To determine if S's role holds
(define (S-INV ci stck)
 (and (andmap (lambda (w) (eq? w 'b)) stck)
      (andmap (lambda (w) (eq? w 'a)) ci)
      (= (%\textcolor{blue}{*}% 2 (length ci)) (length stck))))

;; word (listof %\textcolor{darkgreen}{$\Gamma$}%) %\textcolor{darkgreen}{$\rightarrow$}% Boolean
;;Purpose: To determine if H's role holds
(define (H-INV ci stck)
 (let ([ci-as (filter (lambda (w) (eq? w 'a)) ci)]
       [ci-bs (filter (lambda (w) (eq? w 'b)) ci)])
  (and (equal? ci (append ci-as ci-bs))
       (andmap (lambda (w) (eq? w 'b)) stck)
       (<= (length ci-as) (length (append ci-bs stck)) (%\textcolor{blue}{*}% 2 (length ci-as))))))
\end{lstlisting}
A \flatt{} instructor is likely to immediately notice that there is an incorrect transition into \texttt{H}. The transition that pops a \texttt{b} ought to read a \texttt{b}.

\begin{figure}[t!]
\centering
\includegraphics[scale=0.9]{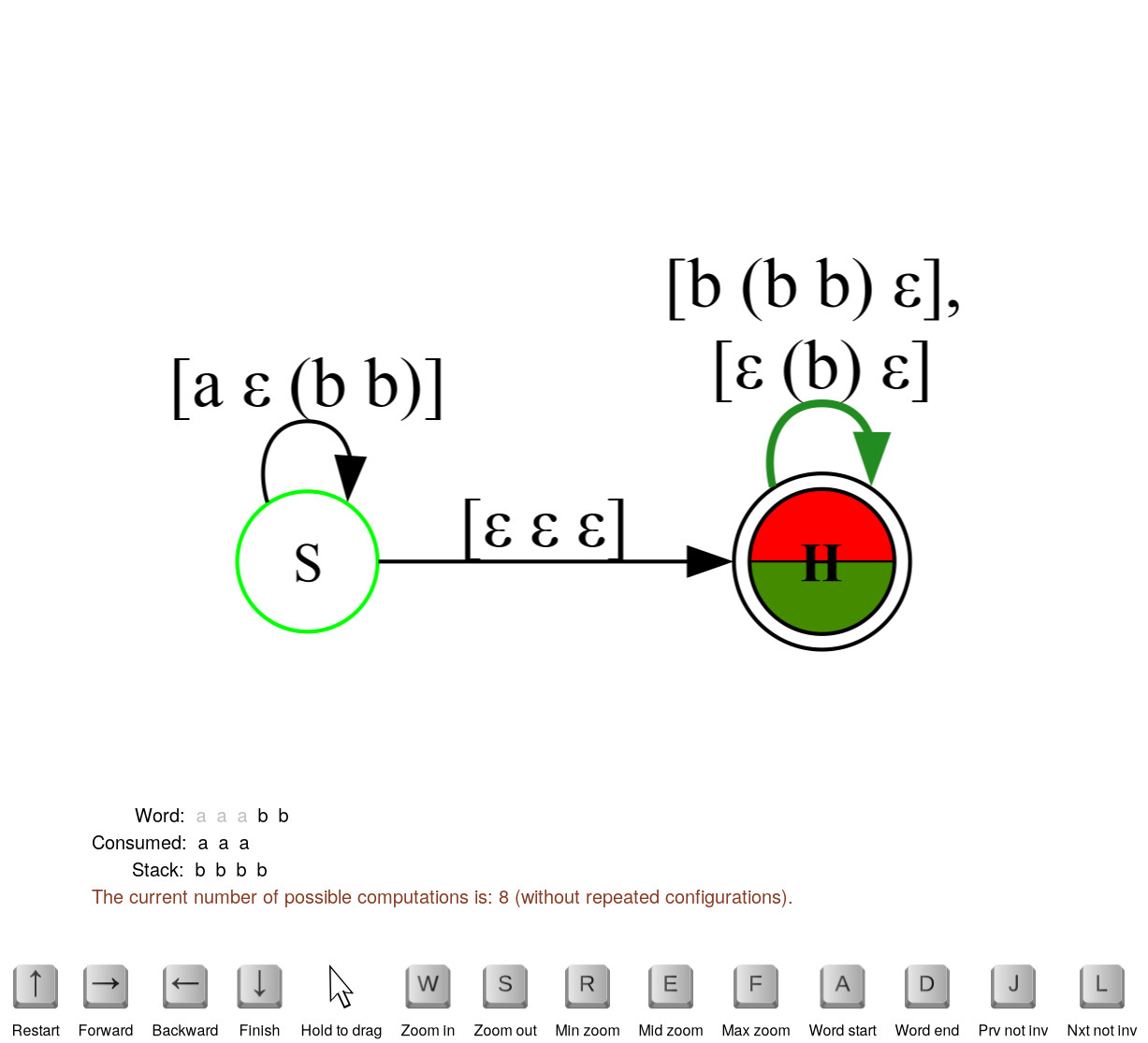}
\caption{\pda{} computations with different invariant values.} \label{bicolor-pda}
\end{figure}

The student uses the visualization tool to discover and correct the bug. Consider the visualization state displayed in \Cref{bicolor-pda}. It illustrates that, after consuming the input word, there are at least two different accepting computations, one in which \texttt{H-INV} holds and one in which \texttt{H-INV} does not hold. This explains the bicolor rendering of \texttt{H}. The student can observe that the given word should not be accepted. Thus, suggesting that there is an error in their transition relation. This leads the student to hone in on the transition \texttt{((H \ep{} (b)) (H \ep{}))} and correct it to \texttt{((H b (b)) (H \ep{}))}. Upon rerunning the visualization tool, the word is rejected and states are not rendered in color because an accepting computation does not exist.

\section{Related Work}
\label{rw}

\begin{figure}[t!]
	\centering
	\begin{subfigure}{.7\textwidth}
		\captionsetup{justification=centering}
		\includegraphics[width=\textwidth]{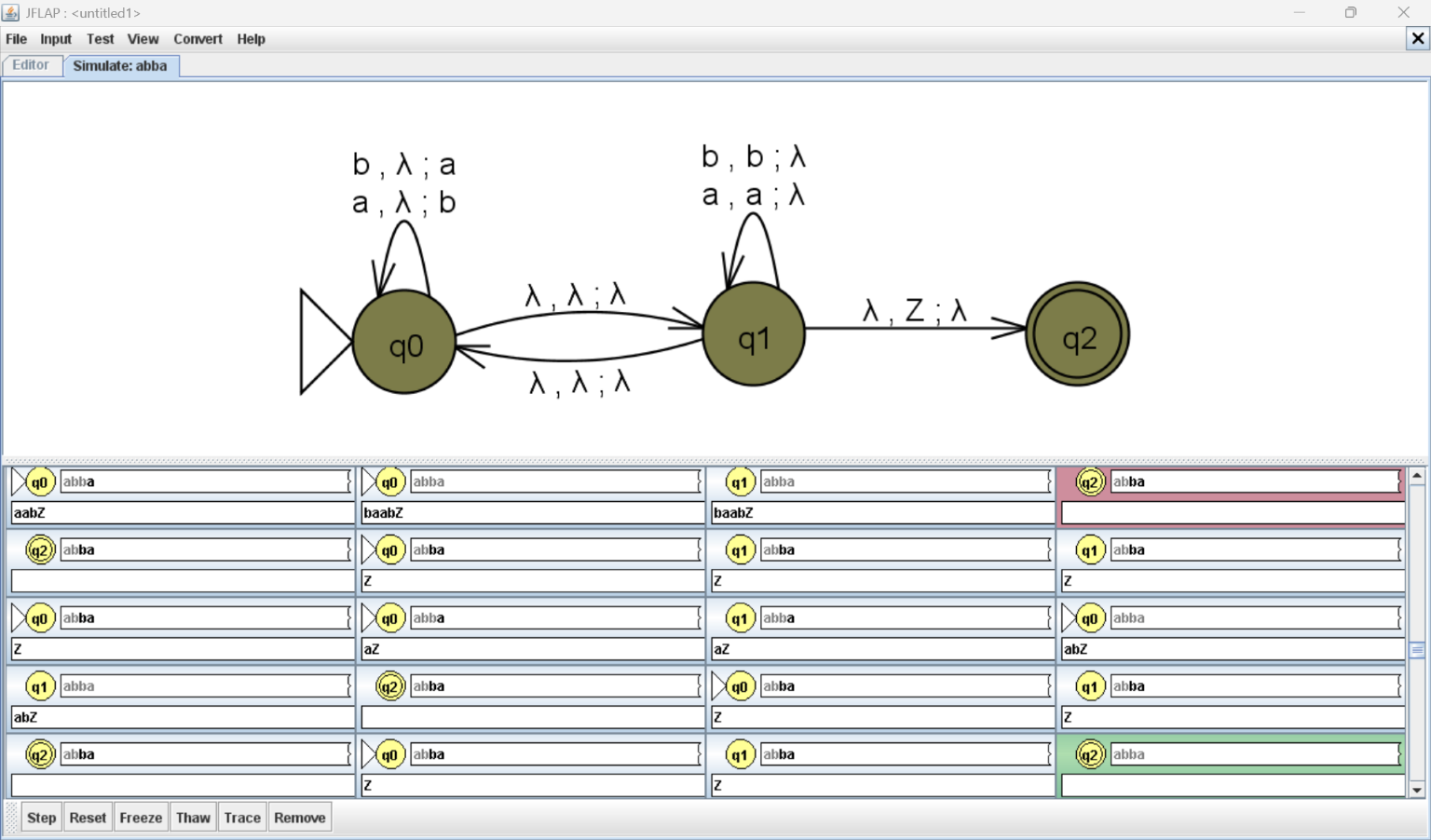}
		\caption{JFLAP's visualization of a \pda{}.}
		\label{rw-1}
	\end{subfigure}
	\begin{subfigure}{.7\textwidth}
		\captionsetup{justification=centering}
		\includegraphics[width=\textwidth]{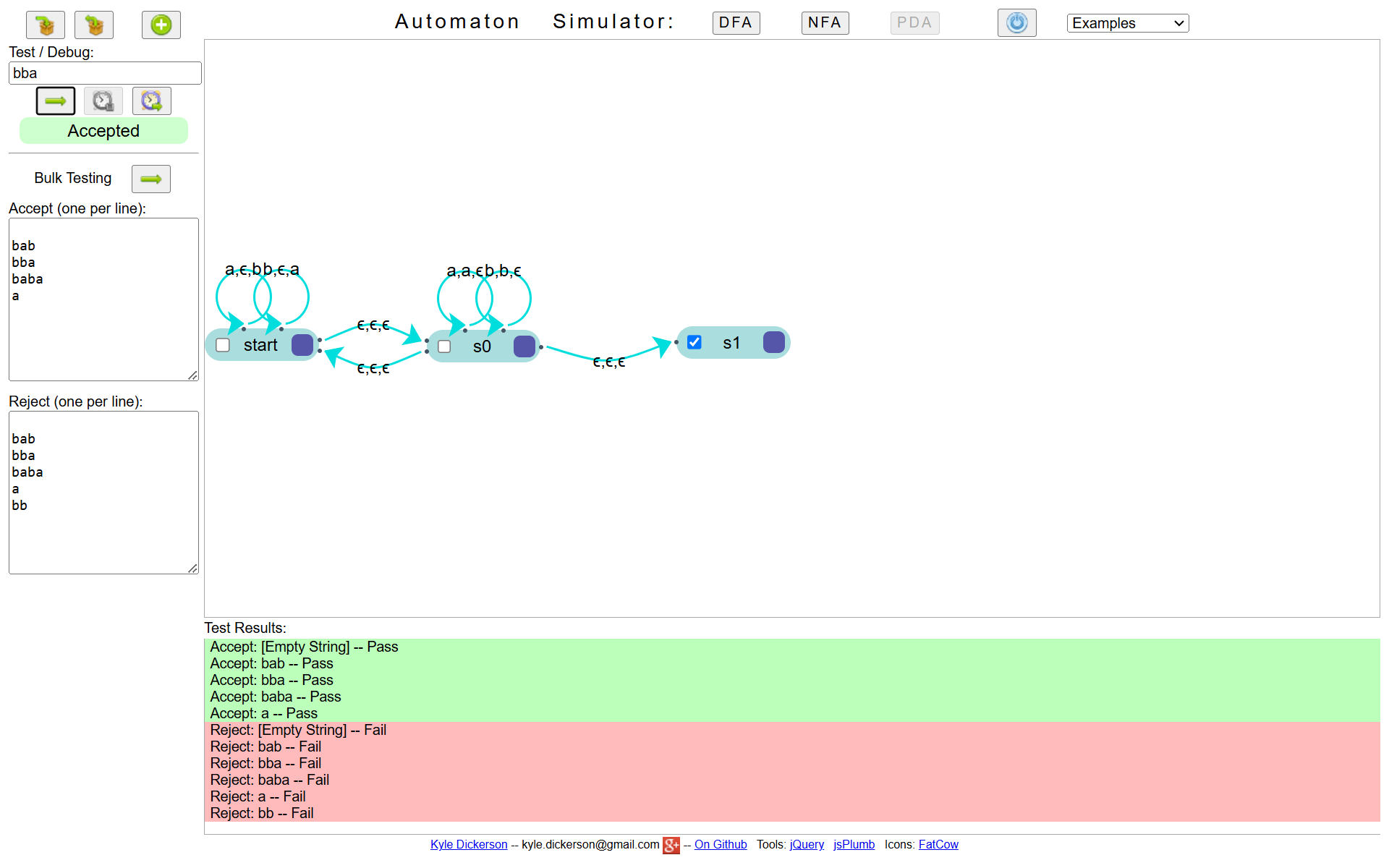}
		\caption{Automaton Simulator's visualization of a \pda{}.}
		\label{rw-2}
	\end{subfigure}
	\label{rw-ss}
	\caption{\pda{} simulations in other tools.}
\end{figure}

\jflap{} is a software tool used to support \flatt{} courses by allowing students to build and test various state machines, including \ndfa{}s and \pda{}s \cite{rodger,rodgerII}. Similar to \fsm{}'s new dynamic visualization tools, \jflap{} traces all possible computations that may be performed by a machine. The rendering of these computations, however, is cumbersome and difficult to understand. Consider, for example, the snapshot of a \pda{} simulation using \jflap{} displayed in \Cref{rw-1}. As the reader can appreciate it is difficult to determine at a glance all the states the machine may be in. In addition, the explicit listing of all possible configurations quickly becomes unwieldy. Unlike \fsm{}, \jflap{} does not allow the user to move backwards or finish the visualization in one step. To further contrast with \fsm{}, \jflap{} allows the user to define how the machine accepts. The user can choose for the machine to accept, like in \fsm{}, when a final state is reached, the input word is fully processed, and the stack is empty. Alternatively, the user can choose for the stack not to be empty upon reaching a final state after fully processing the input word. In addition, \jflap{}'s transition diagrams must be manually constructed which distracts students from their design efforts.

\texttt{Automata Tutor} is an online tool used to support \flatt{} courses by providing grading and personalized feedback for different instructor-created exercises \cite{DAntoni2,DAntoni}. It offers machine simulation using user-provided input words. While a useful tool to provide students with immediate feedback on their designs, \texttt{Automata Tutor} does not show all possible paths a machine can take to consume the input word, thus providing limited insights into how nondeterminism works. In addition, Automata Tutor does not provide students with any guidance on how to verify the design of their machines, other than testing them repeatedly.

\texttt{Automaton Simulator} is a web application tool that is used to help students understand the operational semantics of \ndfa{}s and \pda{}s \cite{ASIM}. Users create such machines, apply such machines to words, and, in debugging mode, observe machine execution on all possible computations. The graphics for transition diagrams become cluttered and confusing because the labels on edges can overlap (see, for example, \Cref{rw-2}). Furthermore, in debugging mode for \pda{}s, the stacks for all computations are displayed, thus, further cluttering the visualization's graphics. In further contrast with \fsm{}, a \pda{} accepts when the input word is fully processed and the machine reaches a final state (i.e., the stack does not have to be empty). Finally, the tool gets trapped in infinite computations.

The \texttt{FSM Simulator} is a web application tool that visually simulates \dfa{}s and \ndfa{}s \cite{FSM-SIM}. Similar to \fsm{}, it allows the user to create, simulate, and visualize their machines. Both \fsm{} and \texttt{FSM Simulator} trace all possible computations, use \gviz{} \cite{gviz2,gviz1} to build transition diagrams, allow the user to move through the simulation in a stepwise manner both forwards and backwards, and draw attention to the states the machine may be in. Unlike \fsm{}, \texttt{FSM Simulator} highlights the node, not the edge leading into the state, to denote what state(s) the machine may be in. Additionally, the tool's instructions are verbose which may place an extraneous cognitive load on students as they have to invest a significant amount of time to learn how to use the tool. Furthermore, the machine's transition diagram, the navigation buttons, and the input word are displayed in two separate sections of the visualization's frame, thus, forcing users to scroll between them as they advance the visualization. Finally, \texttt{FSM Simulator} does not provide support for machine validation.

\section{Concluding Remarks}
\label{concl}

This article presents two novel dynamic visualization tools to help students understand nondeterminism and the operational semantics of \ndfa{}s and \pda{}s. Given a machine and a word, the tools trace all possible computations, without shared configurations, that the machine may perform. Features include moving forward and backwards, jumping to the beginning or end of the visualization, informative messages, zooming and dragging capabilities, using key strokes to interact with the tool, and jumping to the previous or next configuration with a failed state invariant predicate. In addition, both tools allow for students to use an optional argument to add a dead/trap state to help them understand the point at which a computation ends without entirely consuming the input word. The \pda{} visualization tool limits the number of transition steps to prevent getting caught in infinite computations and displays where computations are cut off. The maximum number of steps threshold is adjustable by the programmer. Finally, the user may provide invariant predicates for each state to validate their design role.

Future work includes performing empirical studies to measure student perceptions of the discussed dynamic visualizations. In addition, future work also includes providing improved dynamic visualization support for both Turing machines and multitape Turing machines. This support aims to provide a mechanism to visualize all possible computations that can be performed by such machines as done by the work on finite-state and pushdown automata described in this article.

\balance
\bibliographystyle{eptcs}
\bibliography{sm-viz}
	
\end{document}